\documentclass[12pt]{article}
\usepackage{a4wide}
\usepackage{latexsym}
\usepackage{amsmath}
\usepackage{amsfonts}
\usepackage{amscd}
\usepackage{cite}
\usepackage{graphicx}
\usepackage{float}

\usepackage{pslatex}
\usepackage[latin1]{inputenc}
\usepackage[T1]{fontenc}

\def\bq{\begin{eqnarray}}
\def\eq{\end{eqnarray}}

\begin{document}

\thispagestyle{empty}

\begin{flushright}
  MZ-TH/10-43
\end{flushright}

\vspace{1.5cm}

\begin{center}
  {\Large\bf Jet algorithms in electron-positron annihilation: Perturbative higher order predictions\\
  }
  \vspace{1cm}
  {\large Stefan Weinzierl\\
\vspace{2mm}
      {\small \em Institut f{\"u}r Physik, Universit{\"a}t Mainz,}\\
      {\small \em D - 55099 Mainz, Germany}\\
  } 
\end{center}

\vspace{2cm}

\begin{abstract}\noindent
  {
This article gives results on several jet algorithms in electron-positron annihilation:
Considered are the exclusive sequential recombination algorithms
Durham, Geneva, Jade-E0 and Cambridge, which are typically used in electron-positron annihilation.
In addition also inclusive jet algorithms are studied. 
Results are provided for the inclusive 
sequential recombination algorithms Durham, Aachen and anti-$k_t$,
as well as the infrared-safe cone algorithm SISCone.
The results are obtained in perturbative QCD and are $\mathrm{N}^3$LO for the two-jet rates,
NNLO for the three-jet rates, NLO for the four-jet rates and LO for the five-jet rates.
   }
\end{abstract}

\vspace*{\fill}

\newpage

\section{Introduction}
\label{sec:intro}

Hadronic jets occur in all current high-energy collider experiments.
They can be used to extract fundamental quantities like the strong coupling with high precision,
as it is done in three-jet events in electron-positron annihilation.
Furthermore they often occur in signatures for searches of new physics, 
typical examples are signatures consisting of jets plus missing transverse momentum
for new particle searches at the LHC.
A detailed understanding of jet physics is essential in both cases.

Loosely speaking, a hadronic jet consists of several detected particles in the event, all of them
roughly moving in the same direction.
For quantitative studies this intuitive pictures has to be made more precise.
This is done by a jet algorithm, which groups the observed particles in the event into jets.
There are several jet algorithms on the market, and as a consequence what is finally called a jet depends
on the chosen jet algorithm.
For a better understanding of jet physics it is essential to compare and test the various jet algorithms.

The theoretical description of jet cross sections in high energy 
collider experiments can be done in perturbation theory due to the smallness of the strong coupling.
Recently, the next-to-next-to-leading order (NNLO) predictions for three-jet events in electron-positron annihilation have become
available \cite{GehrmannDeRidder:2007hr,GehrmannDeRidder:2007bj,GehrmannDeRidder:2008ug,GehrmannDeRidder:2009dp,Weinzierl:2008iv,Weinzierl:2009ms,Weinzierl:2009yz}.
It is therefore natural to compare the predictions for different jet algorithms.
In this paper I consider eight different jet algorithms, which can be grouped into two sets of four algorithms each.
In the first group there are four jet algorithms traditionally used in electron-positron annihilation.
These are the algorithms Durham, Geneva, Jade-E0 and Cambridge.
The second class of jet algorithms considered in this paper are inclusive jet algorithms.
Inclusive jet algorithms have their origins in hadron collisions, but can also be considered in 
electron-positron annihilation.
The jet algorithms belonging to this second class are:
the inclusive Durham algorithm, the Aachen algorithm, the anti-$k_t$ algorithm and the SISCone algorithm.

The motivation for this paper is twofold:
First of all three-jet events in electron-positron annihilation can be used to extract the value of the
strong coupling \cite{Dissertori:2007xa,Gehrmann:2008kh,Dissertori:2009ik,Dissertori:2009qa,Bethke:2008hf,Pahl:2008uc,Pahl:2009aa,Becher:2008cf}. 
Here one uses the exclusive jet algorithms from the first group, and -- within this group -- in particular the Durham algorithm.
Although first NNLO results for the exclusive jet rates have already appeared in \cite{GehrmannDeRidder:2008ug,Weinzierl:2008iv,Weinzierl:2009ms},
a careful analysis would need a finer binning and better statistics within the Monte Carlo integration.
Motivated by this request from the experimentalists I therefore provide in this article 
the exclusive jet rates with a fine binning in the jet
resolution parameter and a small Monte Carlo integration error.

Let us now turn to the second motivation. The last years have seen the invention of several new jet algorithm to be used
in hadron-hadron collisions, like the anti-$k_t$-algorithm or the SISCone algorithm.
These algorithms are infrared safe and therefore observables based on these jet algorithms can be calculated within 
perturbation theory.
Although these new algorithms have been invented for hadron-hadron collisions, they can equally well be formulated 
for electron-positron annihilation.
We can therefore study the properties of these new algorithms in the clean environment of electron-positron
annihilation. This may provide valuable information for the behaviour of these algorithms in a hadron-hadron environment.
As a first step in this direction I provide therefore in this article the perturbative predictions for four inclusive
jet algorithms (inclusive Durham, Aachen, anti-$k_t$ and SISCone).

All results in this paper have been obtained with the numerical Monte Carlo program Mercutio2 
\cite{Weinzierl:2009nz,Weinzierl:2008iv,Weinzierl:1999yf}.
This program gives the five-jet rate at leading order (LO), the four-jet rate at 
next-to-leading order (NLO) and the three-jet rate at NNLO.
In addition, the two-jet rate can be deducted at $\mathrm{N}^3$LO from the knowledge of the total hadronic cross section
at order $\alpha_s^3$ and the numbers above.
The numerical Monte Carlo program  relies heavily on research carried out in the past years related to differential NNLO calculations:
Integration techniques for two-loop amplitudes \cite{Gehrmann:1999as,Gehrmann:2000zt,Gehrmann:2001ck,Moch:2001zr,Weinzierl:2002hv,Bierenbaum:2003ud,Weinzierl:2004bn,Moch:2005uc},
the calculation of the relevant tree-, one- and two-loop-amplitudes \cite{Berends:1989yn,Hagiwara:1989pp,Falck:1989uz,Ali:1979rz,Ali:1979wj,Schuler:1987ej,Korner:1990sj,Giele:1992vf,Bern:1997ka,Bern:1997sc,Campbell:1997tv,Glover:1997eh,Garland:2001tf,Garland:2002ak,Moch:2002hm},
routines for the numerical evaluation of polylogarithms \cite{Gehrmann:2001pz,Gehrmann:2001jv,Vollinga:2004sn},
methods to handle infrared singularities \cite{Catani:1997vz,Phaf:2001gc,Catani:2002hc,Kosower:2002su,Kosower:2003cz,Weinzierl:2003fx,Weinzierl:2003ra,Kilgore:2004ty,Frixione:2004is,Gehrmann-DeRidder:2003bm,Gehrmann-DeRidder:2004tv,Gehrmann-DeRidder:2005hi,Gehrmann-DeRidder:2005aw,Gehrmann-DeRidder:2005cm,GehrmannDeRidder:2007jk,Daleo:2009yj,Somogyi:2005xz,Somogyi:2006da,Somogyi:2006db,Catani:2007vq,Somogyi:2008fc,Somogyi:2009ri,Aglietti:2008fe,Bolzoni:2010bt}
and experience from the NNLO calculations of
$e^+ e^- \rightarrow \mbox{2 jets}$ and other processes
\cite{Anastasiou:2004qd,Gehrmann-DeRidder:2004tv,Weinzierl:2006ij,Weinzierl:2006yt,Anastasiou:2002yz,Anastasiou:2003yy,Anastasiou:2003ds,Anastasiou:2004xq,Anastasiou:2005qj,Anastasiou:2007mz,Melnikov:2006di,Anastasiou:2005pn,Catani:2001ic,Catani:2001cr,Grazzini:2008tf,Harlander:2001is,Harlander:2002wh,Harlander:2003ai,Ravindran:2003um,Ravindran:2004mb}.

This article reports the pure QCD perturbative results for the jet rates.
Not included are soft-gluon resummations nor power corrections.
Perturbative electro-weak corrections to three-jet observables have been reported recently 
in \cite{CarloniCalame:2008qn,Denner:2009gx,Denner:2010ia}.

This paper is organised as follows: 
In the next section the set of jet algorithms studied in this paper is described in detail.
In section~\ref{sec:perturbative} the perturbative expansion of the jet rates is reviewed.
The numerical results for the jet rates are given in section~\ref{sec:num}.
Finally, section~\ref{sec:conclusions} contains the conclusions.

\section{Jet algorithms}
\label{sec:def}

A jet algorithm is a procedure for grouping particles into jets.
A jet algorithm may depend on one or more parameters. Usually these parameters define how
``large'' a jet should be.
For a meaningful comparison between experiment and theory a jet algorithm has to be infrared-safe.
We may divide the jet algorithms into two categories: First there are the ``exclusive jet algorithms'', 
where each particle in an event is assigned uniquely to one jet.
The exclusive jet algorithms are predominately used in electron-positron annihilation.
The second class consists of the ``inclusive jet algorithms'', where each particle is either assigned
uniquely to one jet or to no jet at all.
The inclusive jet algorithms are mainly used in hadron-hadron collisions.
It should be mentioned that there are also jet algorithms, which allow the possibility that a particle
is assigned to more than one jet.
Usually one adds then a split-merge procedure, which brings us back to the cases listed above.

Let me start with the specifications for exclusive the jet algorithms. The exclusive jet algorithms are mainly
sequential recombination algorithms and are specified by a clustering procedure.
The clustering procedure usually depends on a resolution variable and a recombination prescription.
In the simplest case the clustering procedure is defined through the following steps:
\begin{enumerate}
\item Define a resolution parameter $y_{cut}$
\item For every pair $(p_k, p_l)$ of final-state particles compute the corresponding
resolution variable $y_{kl}$.
\item If $y_{ij}$ is the smallest value of $y_{kl}$ computed above and
$y_{ij} < y_{cut}$ then combine $(p_i, p_j)$ into a single jet ('pseudo-particle')
with momentum $p_{ij}$ according to a recombination prescription.
\item Repeat until all pairs of objects (particles and/or pseudo-particles)
have $y_{kl} > y_{cut}$.
\end{enumerate}
The various jet algorithms differ in the precise definition of the resolution measure
and the recombination prescription.
The various recombination prescriptions are:
\begin{enumerate}
\item E-scheme:
\bq
E_{ij} = E_i + E_j,
 & &
\vec{p}_{ij} = \vec{p}_i + \vec{p}_j.
\eq
The E-scheme conserves energy and momentum, but for massless particles $i$ and $j$ the recombined four-momentum is not massless.
\item E0-scheme:
\bq
E_{ij} = E_i + E_j,
 & &
\vec{p}_{ij} = \frac{E_i + E_j}{\left| \vec{p}_i + \vec{p}_j \right|} \left( \vec{p}_i + \vec{p}_j \right).
\eq
The E0-scheme conserves energy, but not momentum. For massless particles $i$ and $j$ is the recombined four-momentum again massless. 
\item P-scheme:
\bq
E_{ij} = \frac{\left| \vec{p}_i + \vec{p}_j \right|}{E_i + E_j} \left( E_i + E_j \right) = 
\left| \vec{p}_i + \vec{p}_j \right|,
 & &
\vec{p}_{ij} = \vec{p}_i + \vec{p}_j.
\eq
The P-scheme conserves momentum, but not energy. For massless particles $i$ and $j$ is the recombined four-momentum again massless, 
as in the $E0$-scheme.
\end{enumerate}
For the Durham \cite{Stirling:1991ds}, Geneva \cite{Bethke:1991wk} and Jade-E0 \cite{Bartel:1986ua} jet algorithms the 
resolution variables and the recombination prescriptions are defined as follows:
\begin{align}
 & \mbox{Durham:}     & y_{ij} & = \frac{2 \min(E_i^2, E_j^2) \left( 1 - \cos \theta_{ij} \right)}{Q^2}, & \mbox{E-scheme}, \nonumber \\
 & \mbox{Geneva:}     & y_{ij} & = \frac{8}{9} \; \cdot \; \frac{2 E_i E_j \left( 1 - \cos \theta_{ij} \right)}{\left(E_i + E_j \right)^2}, & \mbox{E-scheme}, \nonumber \\
 & \mbox{Jade-E0:}    & y_{ij} & = \frac{2 E_i E_j \left( 1 - \cos \theta_{ij} \right)}{Q^2}, & \mbox{E0-scheme}, 
\end{align}
where $E_i$ and $E_j$ are the energies of particles $i$ and $j$, and $\theta_{ij}$ is the angle
between $\vec{p}_i$ and $\vec{p}_j$.
$Q$ is the centre-of-mass energy.

The Cambridge algorithm \cite{Dokshitzer:1997in} distinguishes an ordering variable $v_{ij}$ and a resolution variable $y_{ij}$.
The clustering procedure of the Cambridge algorithm is defined as follows:
\begin{enumerate}
\item Select a pair of objects $(p_i,p_j)$ with the minimal value of the ordering
variable $v_{ij}$.
\item If $y_{ij} < y_{cut}$ they are combined, one recomputes the relevant
values of the ordering variable and goes back to the first step.
\item If $y_{ij} \geq y_{cut}$ and $E_i < E_j$ then $i$ is defined as
a resolved jet and deleted from the table.
\item Repeat until only one object is left in the table. This object is
also defined as a jet and clustering is finished.
\end{enumerate} 
As ordering variable
\bq
v_{ij} = 2\left( 1 - \cos \theta_{ij} \right)
\eq
is used. The resolution variable is as in the Durham algorithm
\bq
y_{ij} & = & \frac{2 \min(E_i^2, E_j^2) \left( 1 - \cos \theta_{ij} \right)}{Q^2}
\eq
and the E-scheme is used as recombination prescription.

In this paper we study in addition inclusive jet algorithms. 
The first three inclusive jet algorithms under consideration are again sequential
recombination algorithms.
The clustering procedure is now defined by
\begin{enumerate}
\item Define a resolution parameter $y_{cut}$
\item For every pair $(p_k, p_l)$ of final-state particles compute the corresponding
resolution variable $y_{kl}$.
\item If $y_{ij}$ is the smallest value of $y_{kl}$ computed above and
$y_{ij} < y_{cut}$ then combine $(p_i, p_j)$ into a single jet ('pseudo-particle')
with momentum $p_{ij}$ according to a recombination prescription.
\item Repeat until all pairs of objects (particles and/or pseudo-particles)
have $y_{kl} > y_{cut}$.
\item Objects with $E \ge E_{min}$ are called jets.
\end{enumerate}
This procedure depends on two parameters $y_{cut}$ and $E_{min}$.
Step 5 is new compared to the clustering procedure of the exclusive case: Only pseudo-particles
with an energy larger than $E_{min}$ are called jets.
For the inclusive Durham algorithm, the Aachen algorithm \cite{Wobisch:1998wt}
and the anti-$k_t$ algorithm \cite{Cacciari:2008gp}
the resolution measures and the recombination prescriptions are defined as follows:
\begin{align}
 & \mbox{Durham:}     & y_{ij} & = \frac{2 \min(E_i^2, E_j^2) \left( 1 - \cos \theta_{ij} \right)}{Q^2}, & \mbox{E-scheme}, \nonumber \\
 & \mbox{Aachen:}     & y_{ij} & = \frac{1}{2} \left( 1 - \cos \theta_{ij} \right), & \mbox{E-scheme}, \nonumber \\
 & \mbox{Anti-$k_t$:} & y_{ij} & = \frac{1}{8} Q^2 \min\left(\frac{1}{E_i^2}, \frac{1}{E_j^2}\right) \left( 1 - \cos \theta_{ij} \right), & \mbox{E-scheme},
\end{align}
These resolution measures are all special cases from a family of resolution measures given by
\bq
 y_{ij} & = \frac{1}{2} \left(\frac{Q^2}{4}\right)^{-p} \min\left(E_i^{2p}, E_j^{2p}\right) \left( 1 - \cos \theta_{ij} \right),
\eq
which is parametrised by a variable $p$. 
The resolution measures for the Durham, Aachen and anti-$k_t$ algorithms correspond to $p=1$, $p=0$ and $p=-1$, respectively.
The Aachen algorithm has originally been defined for deep-inelastic scattering, the anti-$k_t$ algorithm has originally 
been defined for hadron-hadron collisions. The definitions above adopt these algorithms to the case of electron-positron annihilation.
The inclusive version of the Durham algorithm differs from the exclusive version of the Durham algorithm by the fact
that at the end of the clustering procedure only pseudo-particles with an energy larger than $E_{min}$ are considered as
(hard) jets. This is due to the additional step 5.
For the Aachen and the anti-$k_t$ algorithms the additional step 5 is essential. 
The resolution measures of these algorithms allow that a single soft parton emitted at large angle to all hard jets
forms a protojet after step 4. This would not be infrared-safe and step 5 removes protojets with an energy smaller
than $E_{min}$.

In addition we consider one further inclusive jet algorithm: 
the infrared-safe cone algorithm SISCone \cite{Salam:2007xv}.
We use the spherical version of this algorithm, which is appropriate 
for electron-positron annihilation.
The algorithm depends on four parameters ($R$, $E_{min}$, $n_{pass}$ and $f$).
The most important parameters are the 
the cone half-opening angle $R$ and the parameter $E_{min}$ defined as above.
For a uniform notation throughout this paper we relate the cone half-opening angle $R$ to a parameter $y_{cut}$ by
\bq
 y_{cut} & = & 1 - \cos R.
\eq
The parameter $n_{pass}$ specifies how often the procedure for finding stable cones is maximally iterated.
The split-merge procedure of this algorithm depends on an overlap parameter $f$. 
In detail the SISCone algorithm is specified as follows:
\begin{enumerate}
\item Put the set of current particles equal to the set of all particles in the event and set $i_{pass}=0$.
\item For the current set of particles find all stable cones with cone half-opening angle $R$. 
\item Each stable cone is added to the list of protojets.
\item Remove all particles that are in stable cones from the list of current particles and increase $i_{pass}$.
\item If $i_{pass}<n_{pass}$ and some new stable cones have been found in this pass, go back to step 2.
\item Run the split-merge procedure with overlap parameter $f$.
\item Objects with $E \ge E_{min}$ are called jets.
\end{enumerate}
A set of particles defines a cone axis, which is given as the sum of the momenta of all particles in the set.
A cone is called stable for the cone half-opening angle $R$, 
if all particles defining the cone axis have an angle smaller than $R$ to the cone
axis and if all particles not belonging to the cone have an angle larger than $R$ with respect to the cone axis.
The angle $\theta$ between a particle with three-momentum $\vec{p}_j$ and a cone axis given by $\vec{p}_{cone}$ is given
by
\bq
 \cos \theta & = & \frac{\vec{p}_j \cdot \vec{p}_{cone}}{\left|\vec{p}_j \right| \left| \vec{p}_{cone}\right|}.
\eq
The four-momentum of a protojet is the sum of the four-momenta of the particles in the protojet.
This corresponds to the E-scheme.
The split-merge procedure requires an infrared-safe ordering variable for the protojets. 
For events in electron-positron annihilation this ordering variable is denoted $\tilde{E}$
and is given for a protojet $I$ with three-momentum $\vec{p}_I$ by 
\bq
 \tilde{E}(I) & = & \sum\limits_{k \in I} E_k \left[ 1 + \sin^2\theta\left(\vec{p}_k,\vec{p}_I\right) \right].
\eq
The sum is over all particles in the protojet.
The split-merge procedure is defined as follows:
\begin{enumerate}
\item Find the protojet $I$ with the highest $\tilde{E}(I)$.
\item Among the remaining protojets find the one ($J$) with highest $\tilde{E}(J)$ that overlaps with $J$.
\item If there is such an overlapping jet then compute the sum of the energies of the particles shared by $I$ and $J$:
\bq 
 E_{shared} & = & \sum\limits_{k\in (I \cap J)} E_k.
\eq
\begin{enumerate}
\item If $E_{shared} < f E(J)$ assign each particle that is shared between the two protojets to the protojet whose
axis is closest. Recalculate the momenta of the protojets.
\item If $E_{shared} \ge f E(J)$ merge the two protojets into a single new protojet and remove the two original
ones.
\end{enumerate}
\item Otherwise, if no overlapping jet exists, then add $I$ to the list of jets and remove it from the list
of protojets.
\item As long as there are protojets left, go back to step 1.
\end{enumerate}
It should be noted that the SISCone algorithm described here is the one adapted to electron-positron
annihilation. This version uses opening angles as a distance measure in contrast to the distance
measure based on rapidity and azimuthal angle which is typically used in hadron-hadron collisions.
Furthermore, the energy $E$ and the ordering quantity $\tilde{E}$ are used 
in the split-merge procedure instead of $p_t$ and $\tilde{p}_t$.

In this article we will always keep the default value of infinity for the parameter $n_{pass}$.
This ensures that all particles are associated to protojets.
Furthermore we will also always keep the default value of $f=1/2$ for the overlap parameter.
The original implementation of \cite{Salam:2007xv} foresees the possibility that in step 1 of the split-merge procedure 
protojets with an energy smaller than a threshold $E_{threshold}$ are immediately discarded.
We can set the value $E_{threshold}$ to zero, since we keep in step 7 at the end of the main algorithm 
only jets with an energy larger than $E_{min}$.

The inclusive jet algorithms all depend on a parameter $E_{min}$. It is convenient to introduce a dimensionless
quantity $\eta_{min}$, related to $E_{min}$ by
\bq
 E_{min} & = & \eta_{min} \sqrt{Q^2},
\eq
where $\sqrt{Q^2}$ is the centre-of-mass energy.
Throughout this paper we use the value
\bq
 \eta_{min} & = & 0.077.
\eq
This value corresponds to $E_{min}=7\;\mbox{GeV}$ for $\sqrt{Q^2}=m_Z$ and is motivated by the value
used by the OPAL collaboration \cite{Akers:1994wj}.

\section{Perturbative expansion}
\label{sec:perturbative}

The production rate for $n$-jet events in electron-positron annihilation is given as
the ratio of the cross section for $n$-jet events divided by the total hadronic cross section
\bq
 R_n(\mu) & = &  \frac{\sigma_{n-jet}(\mu)}{\sigma_{tot}(\mu)}.
\eq
The arbitrary renormalisation scale is denoted by $\mu$.
The production rates can be calculated within perturbation theory.
Assuming that the jet algorithm does not classify any event as a one-jet or zero-jet event, we have the 
perturbative expansions
\bq
\begin{aligned}
 R_2(\mu) & = 
  & 1 + \frac{\alpha_s(\mu)}{2\pi} \bar{A}_2(\mu)
    + & \left( \frac{\alpha_s(\mu)}{2\pi} \right)^2 \bar{B}_2(\mu)
    & + \left( \frac{\alpha_s(\mu)}{2\pi} \right)^3 \bar{C}_2(\mu)
    & + {\cal O}(\alpha_s^4), 
 \nonumber \\
 R_3(\mu) & =  
    &   \frac{\alpha_s(\mu)}{2\pi} \bar{A}_3(\mu)
    + & \left( \frac{\alpha_s(\mu)}{2\pi} \right)^2 \bar{B}_3(\mu)
    & + \left( \frac{\alpha_s(\mu)}{2\pi} \right)^3 \bar{C}_3(\mu)
    & + {\cal O}(\alpha_s^4), 
 \nonumber \\
 R_4(\mu) & =  
    &   
    & \left( \frac{\alpha_s(\mu)}{2\pi} \right)^2 \bar{B}_4(\mu)
    & + \left( \frac{\alpha_s(\mu)}{2\pi} \right)^3 \bar{C}_4(\mu)
    & + {\cal O}(\alpha_s^4), 
 \nonumber \\
 R_5(\mu) & =  
    & 
    & 
    & \left( \frac{\alpha_s(\mu)}{2\pi} \right)^3 \bar{C}_5(\mu)
    & + {\cal O}(\alpha_s^4). 
 \\
\end{aligned}
\eq
If the jet algorithms allows the possibility that an event is classified as a one-jet or zero-jet event, 
we can keep the notation as above, but have to interpret $R_2$ as the production rate for events with no more than two jets.
This occurs for example in the SISCone algorithm, where a tiny fraction of events are classified as one-jet events due to the
split-merge procedure.

In practise the numerical program computes the quantities
\bq
\begin{aligned}
 \frac{\sigma_{3-jet}(\mu)}{\sigma_{0}(\mu)} & =  
    &   \frac{\alpha_s(\mu)}{2\pi} A_3(\mu)
    + & \left( \frac{\alpha_s(\mu)}{2\pi} \right)^2 B_3(\mu)
    & + \left( \frac{\alpha_s(\mu)}{2\pi} \right)^3 C_3(\mu)
    & + {\cal O}(\alpha_s^4), 
 \nonumber \\
 \frac{\sigma_{4-jet}(\mu)}{\sigma_{0}(\mu)} & =  
    &   
    & \left( \frac{\alpha_s(\mu)}{2\pi} \right)^2 B_4(\mu)
    & + \left( \frac{\alpha_s(\mu)}{2\pi} \right)^3 C_4(\mu)
    & + {\cal O}(\alpha_s^4), 
 \nonumber \\
 \frac{\sigma_{5-jet}(\mu)}{\sigma_{0}(\mu)} & =  
    & 
    & 
    & \left( \frac{\alpha_s(\mu)}{2\pi} \right)^3 C_5(\mu)
    & + {\cal O}(\alpha_s^4),
 \\
\end{aligned}
\eq
normalised to $\sigma_0$, which is the LO cross section for $e^+ e^- \rightarrow \mbox{hadrons}$,
instead of the normalisation to $\sigma_{tot}$.
The corresponding coefficients $A_2(\mu)$, $B_2(\mu)$ and $C_2(\mu)$ for the two-jet rate  
\bq
 \frac{\sigma_{2-jet}(\mu)}{\sigma_{0}(\mu)} & = &
    1 + \frac{\alpha_s(\mu)}{2\pi} A_2(\mu)
    + \left( \frac{\alpha_s(\mu)}{2\pi} \right)^2 B_2(\mu)
    + \left( \frac{\alpha_s(\mu)}{2\pi} \right)^3 C_2(\mu)
    + {\cal O}(\alpha_s^4)
\eq
are obtained from the coefficients of three-, four- and five-jet rates and the known perturbative expansion
of the total hadronic cross section $\sigma_{tot}$
\bq
 \sigma_{tot}(\mu) =  
 \sigma_0(\mu) \left( 1 + \frac{\alpha_s(\mu)}{2\pi} A_{tot}(\mu)
                   + \left( \frac{\alpha_s(\mu)}{2\pi} \right)^2 B_{tot}(\mu)
                   + \left( \frac{\alpha_s(\mu)}{2\pi} \right)^3 C_{tot}(\mu)
                   + {\cal O}(\alpha_s^4) \right).
\eq
We have
\bq
 A_2 & = & A_{tot} - A_3,
 \nonumber \\
 B_2 & = & B_{tot} - B_3 - B_4,
 \nonumber \\
 C_2 & = & C_{tot} - C_3 - C_4 - C_5.
\eq
There is a simple relation between the coefficients $A_n$, $B_n$ and $C_n$
and the coefficients $\bar{A}_n$, $\bar{B}_n$ and $\bar{C}_n$:
\bq
 \bar{A}_n & = & A_n - \delta_{n,2} A_{tot},
 \nonumber \\
 \bar{B}_n & = & B_n - A_{tot} A_n - \delta_{n,2} \left( B_{tot} - A_{tot}^2 \right),
 \nonumber \\
 \bar{C}_n & = & C_n - A_{tot} B_n - \left( B_{tot} - A_{tot}^2 \right) A_n - \delta_{n,2} \left( C_{tot} - 2 A_{tot} B_{tot} + A_{tot}^3 \right).
\eq
It is sufficient to calculate the functions $\bar{A}_{\cal O}$, $\bar{B}_{\cal O}$ and $\bar{C}_{\cal O}$
for a fixed renormalisation scale $\mu_0$, which can be taken conveniently to be equal to the
centre-of-mass energy: $\mu_0=Q$.
For this scale choice the coefficients of the perturbative expansion of the total hadronic cross section are given by
\cite{Gorishnii:1991vf,Surguladze:1990tg}:
\bq
\label{coeff_total_cross_section}
 A_{tot} & = & \frac{3}{2} C_F,
 \nonumber \\
 B_{tot} & = &
 \frac{1}{4} \left[ - \frac{3}{2} C_F^2 + C_F C_A \left( \frac{123}{2} - 44 \zeta_3 \right)
                    + C_F T_R N_f \left( -22 + 16 \zeta_3 \right) \right],
 \nonumber \\
 C_{tot} & = &
 \frac{1}{8} \left[
   - \frac{69}{2} C_F^3
   + C_F^2 C_A \left( -127 - 572 \zeta_3 + 880 \zeta_5 \right)
   + C_F C_A^2 \left( \frac{90445}{54} - \frac{10948}{9} \zeta_3 - \frac{440}{3} \zeta_5 \right)
 \right. \nonumber \\
 & & \left.
   + C_F^2 T_R N_f  \left( -29+304\zeta_3 -320 \zeta_5 \right)
   + C_F C_A T_R N_f \left( -\frac{31040}{27} + \frac{7168}{9} \zeta_3 + \frac{160}{3} \zeta_5 \right)
 \right. \nonumber \\
 & & \left.
   + C_F T_R^2 N_f^2 \left( \frac{4832}{27} - \frac{1216}{9} \zeta_3 \right)
   - \pi^2 C_F \left( \frac{11}{3} C_A - \frac{4}{3} T_R N_f \right)^2
 \right].
\eq
The colour factors are defined as usual by
\bq
 C_A = N_c,
 \;\;\;
 C_F = \frac{N_c^2-1}{2 N_c},
 \;\;\;
 T_R = \frac{1}{2}.
\eq
$N_c$ denotes the number of colours and $N_f$ the number of light quark flavours.
In eq.~(\ref{coeff_total_cross_section}) there are in addition singlet contributions to the coefficient
$C_{tot}$, which arise from interference terms of amplitudes, where
the electro-weak boson couples to two different fermion lines.
These contributions are not shown in eq.~(\ref{coeff_total_cross_section}) and neglected throughout this paper.
At present, these singlet contributions are known at order $\alpha_s^3$ for the total hadronic cross section and the 
four- and five-jet cross sections. They are not known for the three-jet cross section, but can be expected to be numerically
small \cite{Dixon:1997th,vanderBij:1988ac,Garland:2002ak}.

The scale variation can be restored from the renormalisation group equation
\bq
\label{RGE_alpha_s}
 \mu^2 \frac{d}{d\mu^2} \left( \frac{\alpha_S}{2\pi} \right)
 & = & 
 - \frac{1}{2} \beta_0 \left( \frac{\alpha_S}{2\pi} \right)^2
 - \frac{1}{4} \beta_1 \left( \frac{\alpha_S}{2\pi} \right)^3
 - \frac{1}{8} \beta_2 \left( \frac{\alpha_S}{2\pi} \right)^4
 + {\cal O}(\alpha_s^5),
 \\
 \beta_0 & = & \frac{11}{3} C_A - \frac{4}{3} T_R N_f,
 \nonumber \\
 \beta_1 & = & \frac{34}{3} C_A^2 - 4 \left( \frac{5}{3} C_A + C_F \right) T_R N_f,
 \nonumber \\
 \beta_2 & = & \frac{2857}{54} C_A^3 - \left( \frac{1415}{27} C_A^2 + \frac{205}{9} C_A C_F - 2 C_F^2 \right) T_R N_f
             + \left( \frac{158}{27} C_A + \frac{44}{9} C_F \right) T_R^2 N_f^2.
 \nonumber
\eq
The values of the coefficients $\bar{A}_n$, $\bar{B}_n$ and $\bar{C}_n$
at a scale $\mu$ are then obtained from the ones at the scale $\mu_0$ by
\bq
 \bar{A}_n(\mu) & = & \bar{A}_n(\mu_0),
 \nonumber \\
 \bar{B}_n(\mu) & = & \bar{B}_n(\mu_0) + \frac{1}{2} \beta_0 \ln\left(\frac{\mu^2}{\mu_0^2}\right) \bar{A}_n(\mu_0),
 \nonumber \\
 \bar{C}_n(\mu) & = & \bar{C}_n(\mu_0) + \beta_0 \ln\left(\frac{\mu^2}{\mu_0^2}\right) \bar{B}_n(\mu_0)
                           + \frac{1}{4} \left[ \beta_1 + \beta_0^2 \ln\left(\frac{\mu^2}{\mu_0^2}\right) \right] \ln\left(\frac{\mu^2}{\mu_0^2}\right) \bar{A}_n(\mu_0).
\eq
Finally, an approximate solution of eq.~(\ref{RGE_alpha_s}) for $\alpha_s$ is given by
\bq
 \frac{\alpha_s(\mu)}{2\pi}
 & = &
 \frac{2}{\beta_0 L}
 \left\{
         1 - \frac{\beta_1}{\beta_0^2} \frac{\ln L}{L}
         + \frac{1}{\beta_0^4 L^2} \left[ \beta_1^2 \left( \ln^2L - \ln L - 1\right) + \beta_0 \beta_2 \right]
 \right\},
\eq
where $L=\ln(\mu^2/\Lambda^2)$. This solution is appropriate for NNLO. 
The lower order solutions are obtained by dropping the corresponding higher order terms in $1/L$.
Note that in addition the scale parameter $\Lambda$ has to be adjusted.
We use the LO formula for $\alpha_s$ in the five-jet rates, the NLO formula for $\alpha_s$ in the four-jet rates
and the NNLO formula for $\alpha_s$ in the three- and two-jet rates.
Note that it is consistent to use the NNLO formula for $\alpha_s$ in the $\mathrm{N}^3$LO calculation for the two-jet rate.
The leading order prediction for the two-jet rate is independent of $\alpha_s$.
Using the $\mathrm{N}^3$LO formula for $\alpha_s$ would not improve the theoretical prediction.
It would merely include some -- but not all -- higher order terms.

\section{Numerical results}
\label{sec:num}

\begin{figure}[p]
\begin{center}
\includegraphics[bb= 125 460 490 710,width=0.8\textwidth]{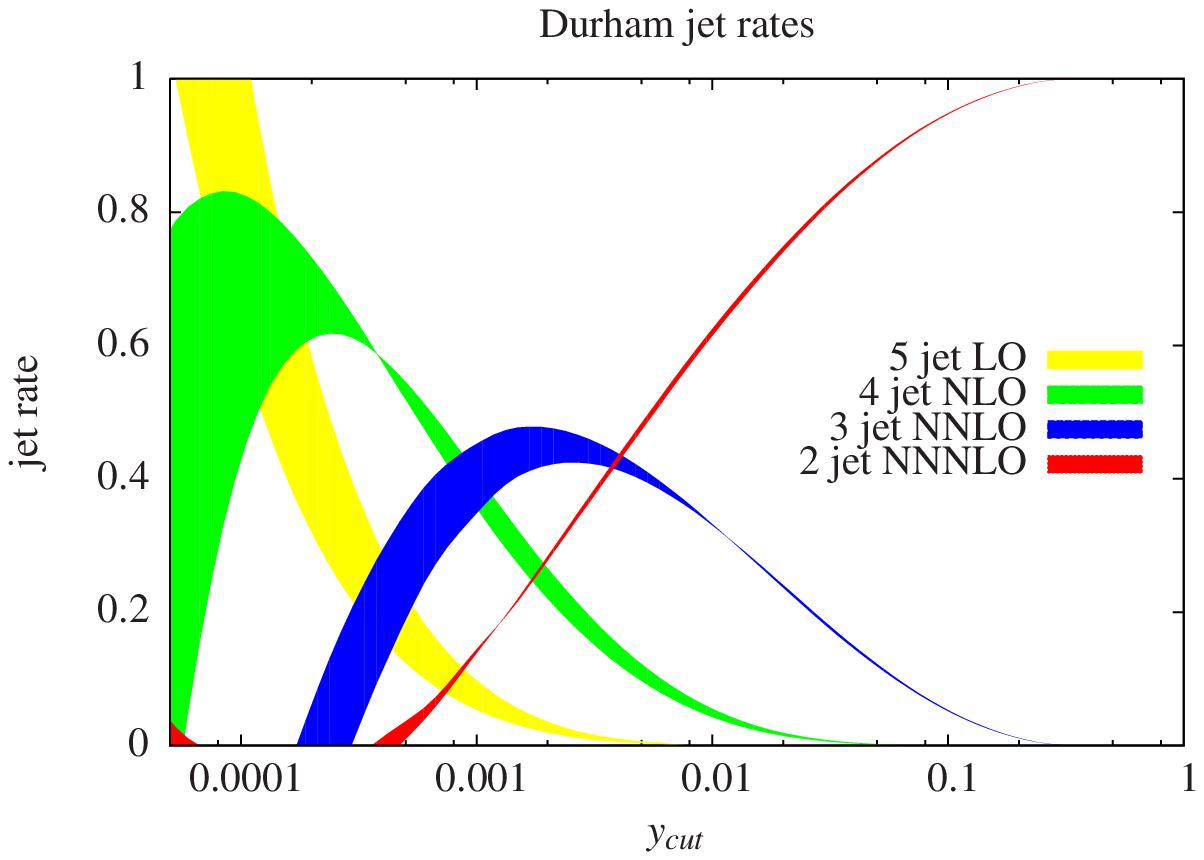}
\includegraphics[bb= 125 460 490 710,width=0.8\textwidth]{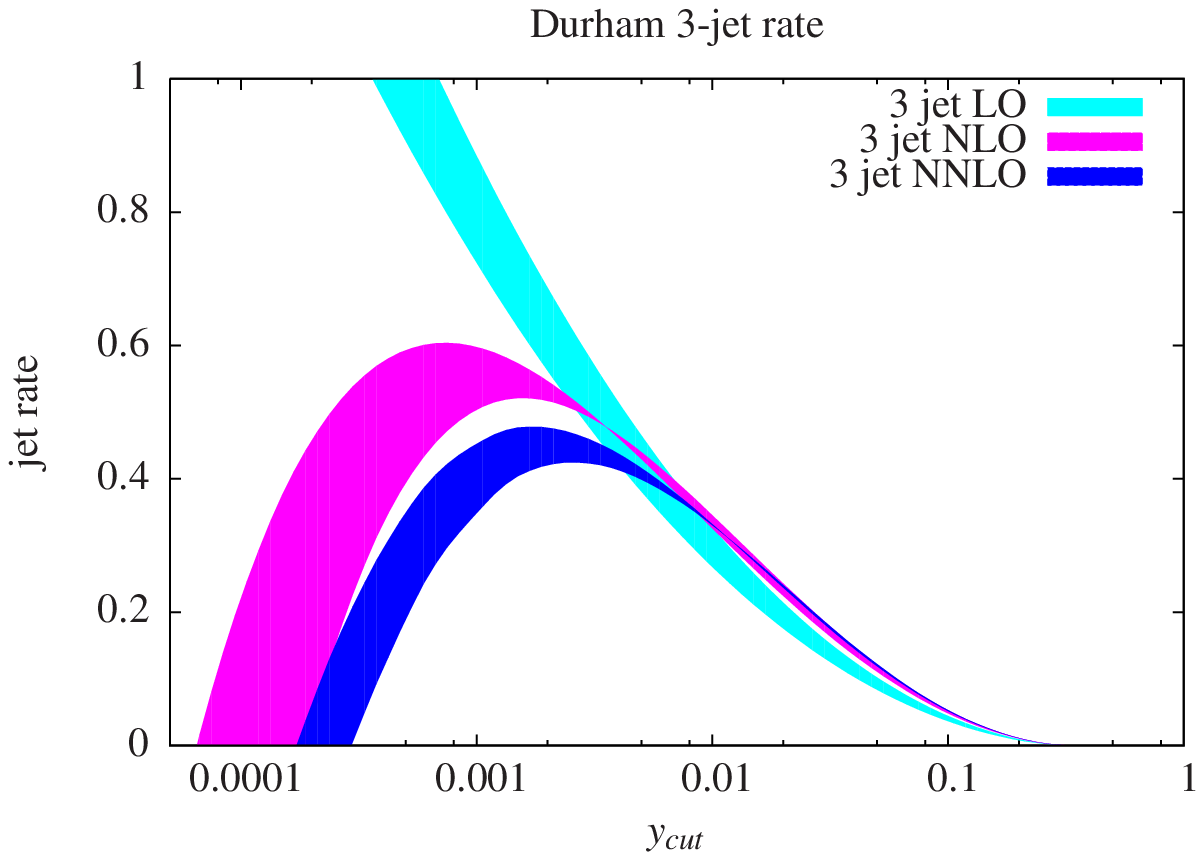}
\end{center}
\caption{
The upper plot shows the jet rates for the Durham algorithm at order $\alpha_s^3$.
The lower plot shows the three-jet rate for the Durham algorithm at LO, NLO and NNLO.
All plots are done with $\sqrt{Q^2}=m_Z$ and $\alpha_s(m_Z)=0.118$.
The bands give the range for the theoretical prediction obtained from varying the renormalisation scale
from $\mu=m_Z/2$ to $\mu=2 m_Z$.
}
\label{fig_durham}
\end{figure}

\begin{figure}[p]
\begin{center}
\includegraphics[bb= 125 460 490 710,width=0.8\textwidth]{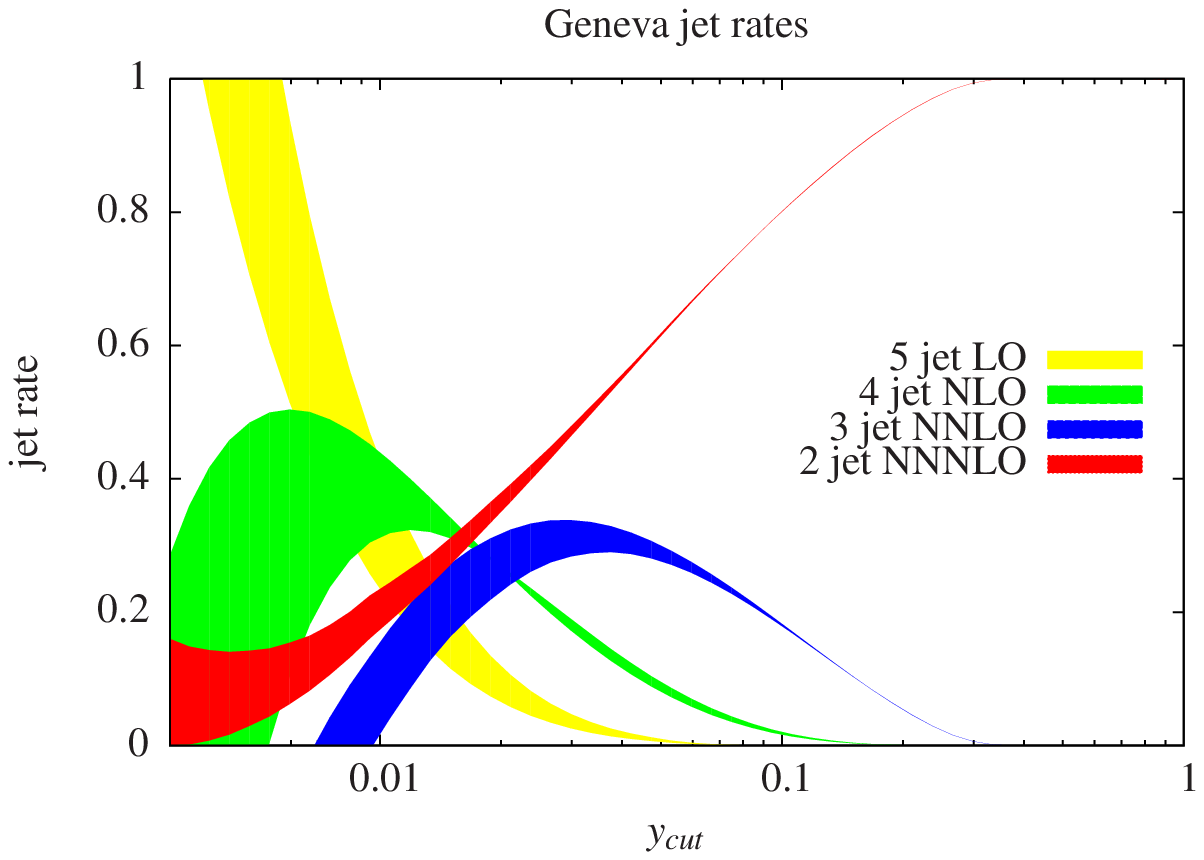}
\includegraphics[bb= 125 460 490 710,width=0.8\textwidth]{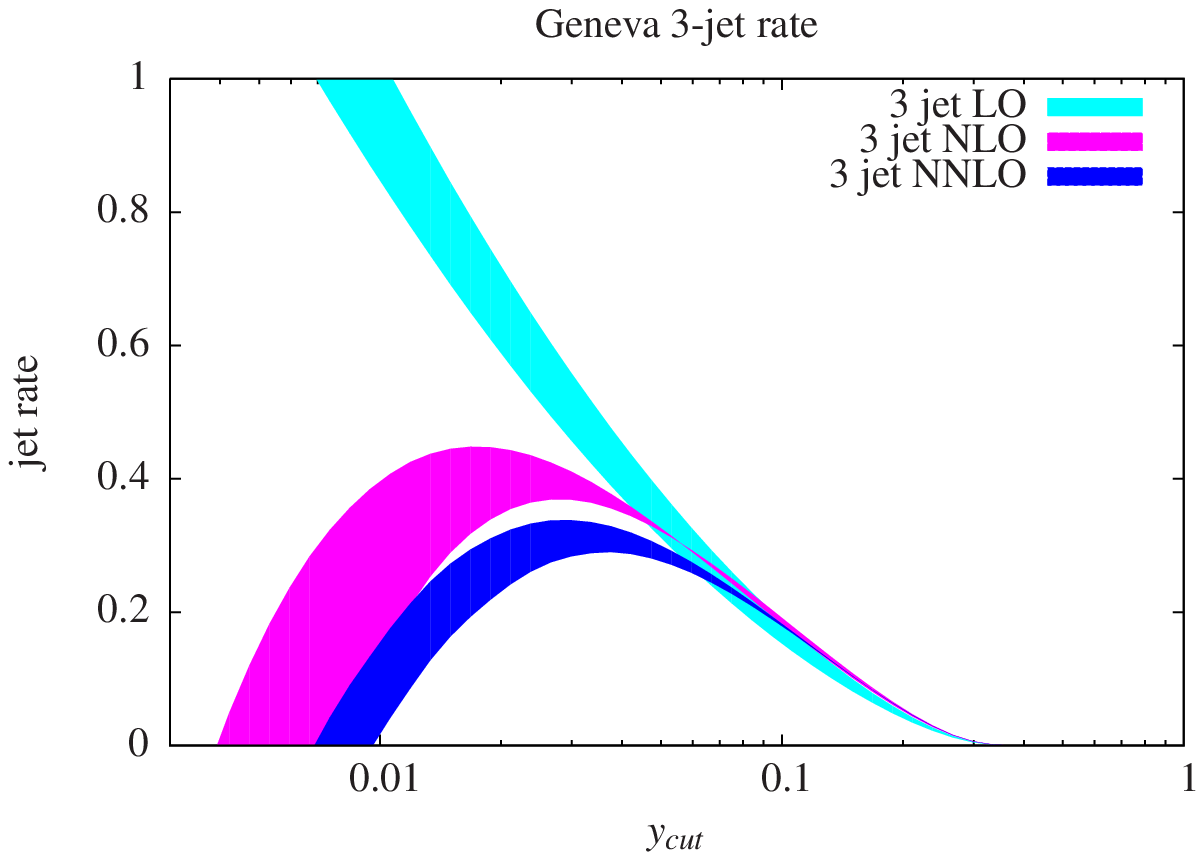}
\end{center}
\caption{
The upper plot shows the jet rates for the Geneva algorithm at order $\alpha_s^3$.
The lower plot shows the three-jet rate for the Geneva algorithm at LO, NLO and NNLO.
All plots are done with $\sqrt{Q^2}=m_Z$ and $\alpha_s(m_Z)=0.118$.
The bands give the range for the theoretical prediction obtained from varying the renormalisation scale
from $\mu=m_Z/2$ to $\mu=2 m_Z$.
}
\label{fig_geneva}
\end{figure}

\begin{figure}[p]
\begin{center}
\includegraphics[bb= 125 460 490 710,width=0.8\textwidth]{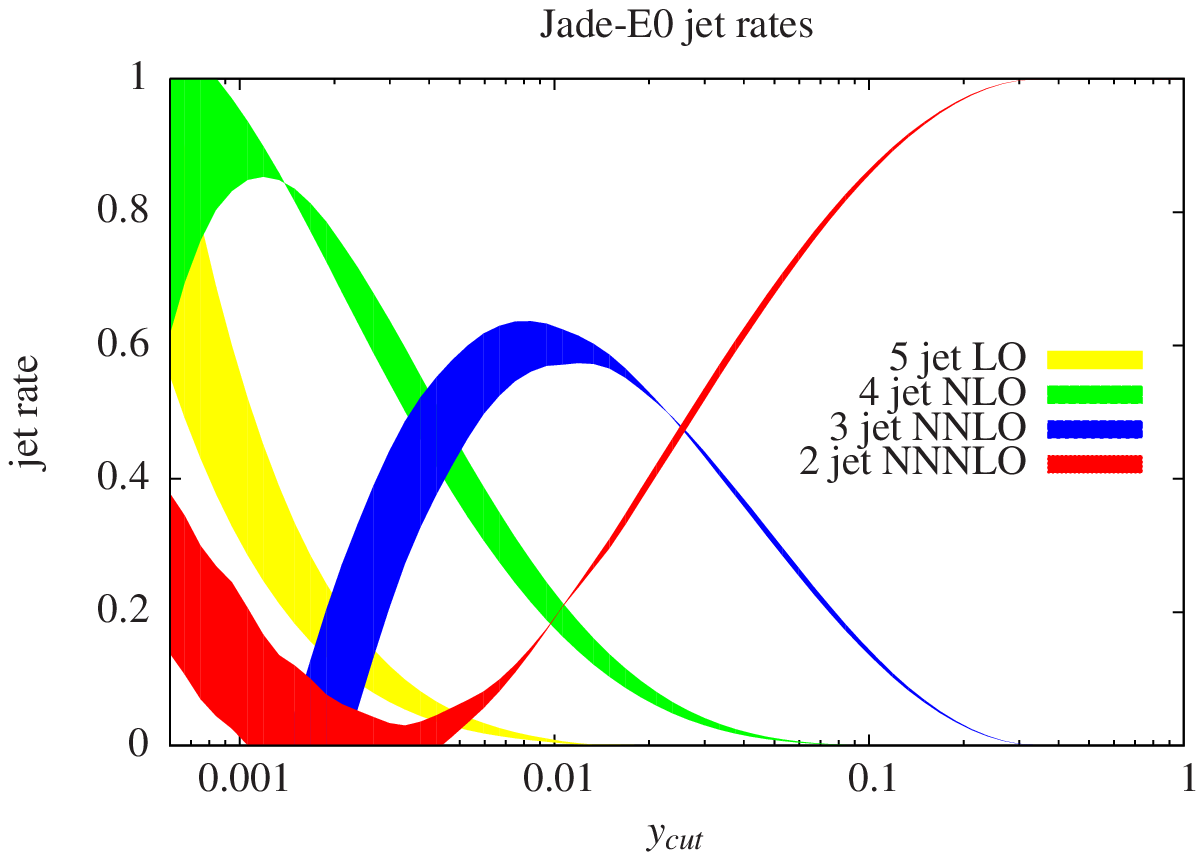}
\includegraphics[bb= 125 460 490 710,width=0.8\textwidth]{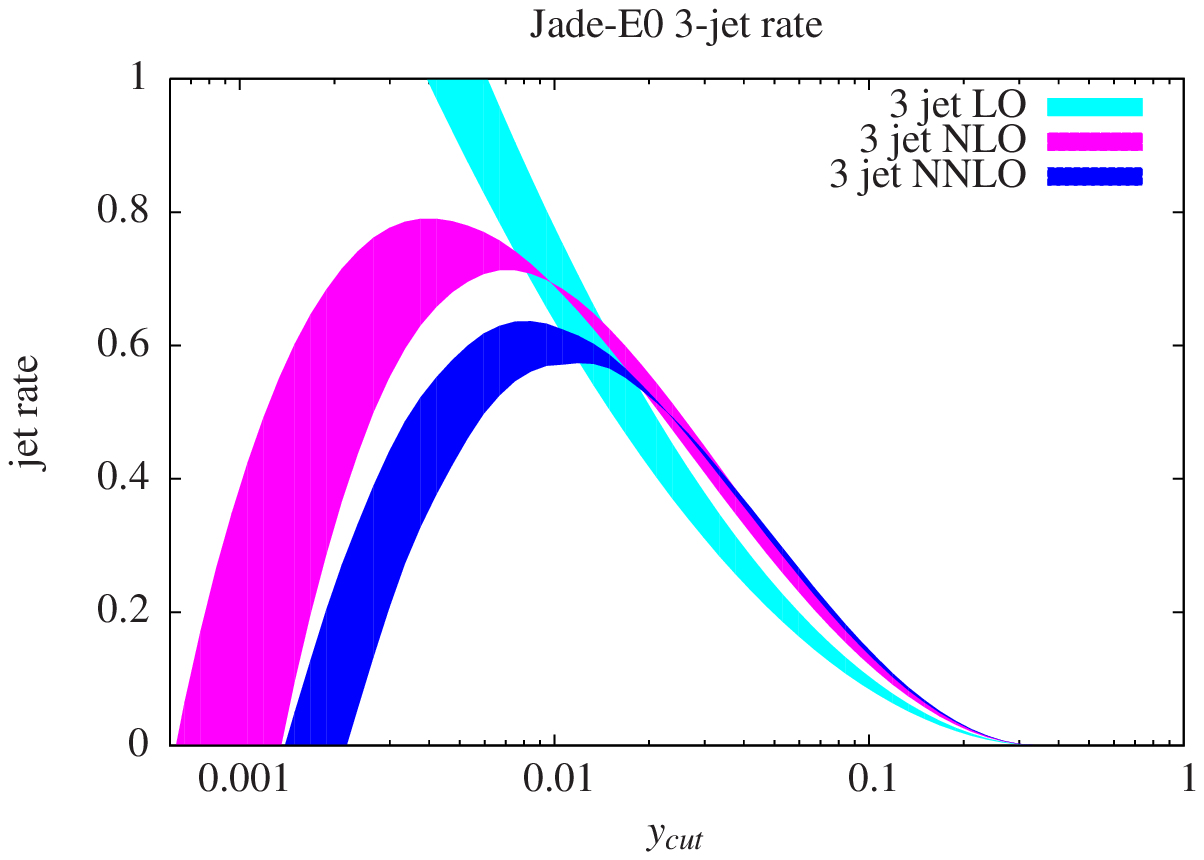}
\end{center}
\caption{
The upper plot shows the jet rates for the Jade-E0 algorithm at order $\alpha_s^3$.
The lower plot shows the three-jet rate for the Jade-E0  algorithm at LO, NLO and NNLO.
All plots are done with $\sqrt{Q^2}=m_Z$ and $\alpha_s(m_Z)=0.118$.
The bands give the range for the theoretical prediction obtained from varying the renormalisation scale
from $\mu=m_Z/2$ to $\mu=2 m_Z$.
}
\label{fig_jadeE0}
\end{figure}

\begin{figure}[p]
\begin{center}
\includegraphics[bb= 125 460 490 710,width=0.8\textwidth]{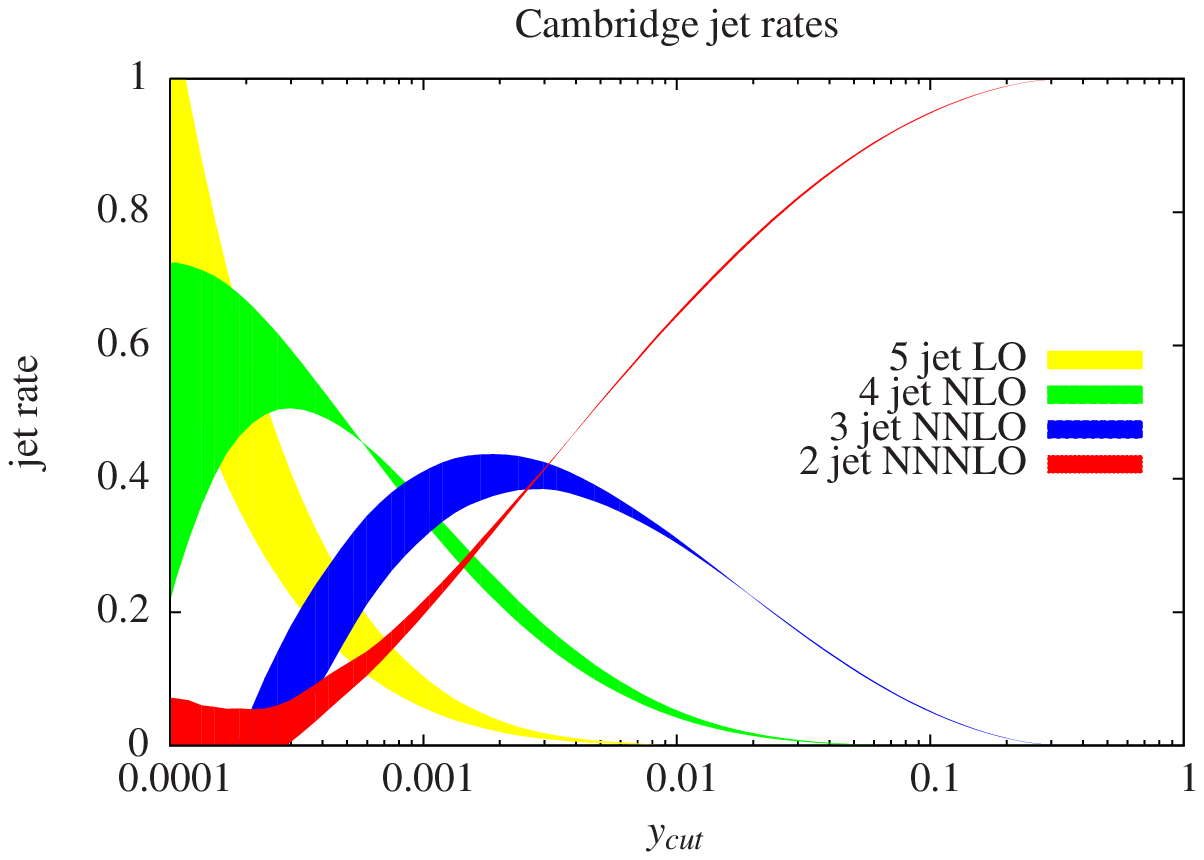}
\includegraphics[bb= 125 460 490 710,width=0.8\textwidth]{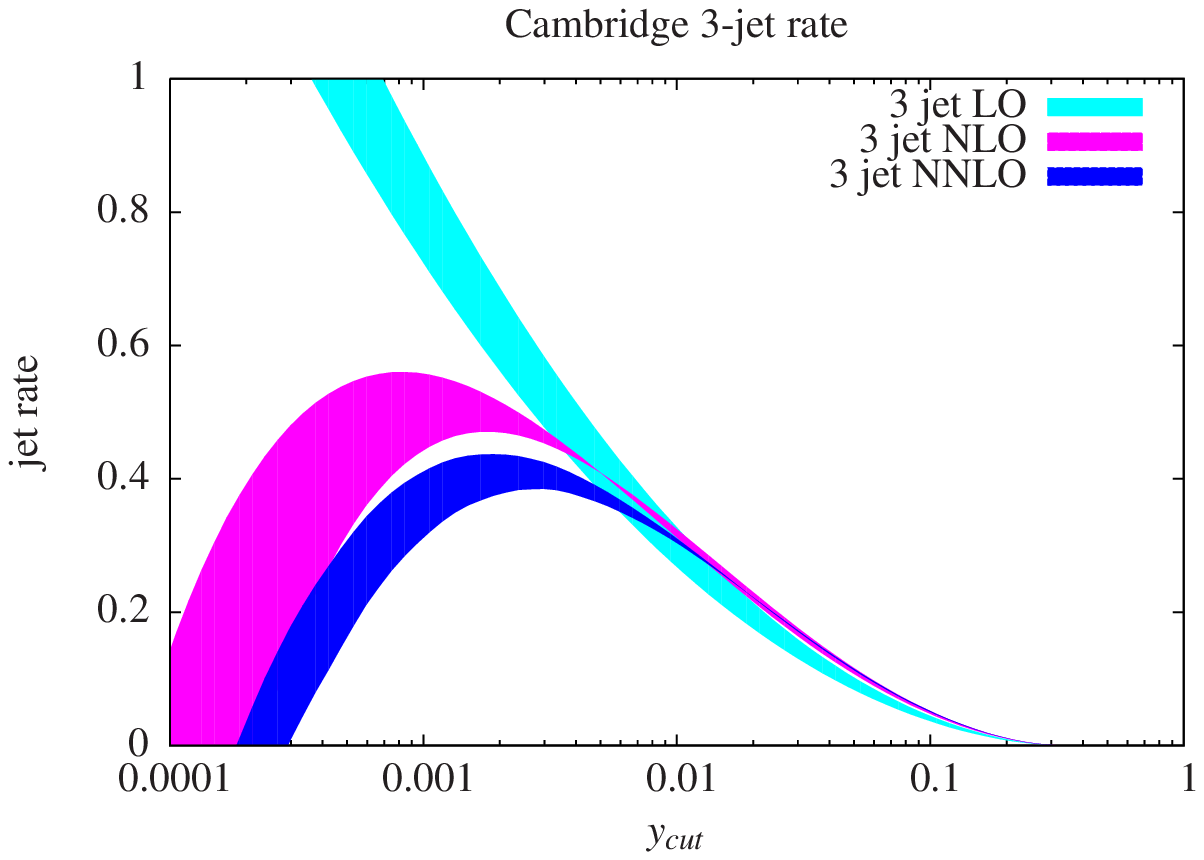}
\end{center}
\caption{
The upper plot shows the jet rates for the Cambridge algorithm at order $\alpha_s^3$.
The lower plot shows the three-jet rate for the Cambridge algorithm at LO, NLO and NNLO.
All plots are done with $\sqrt{Q^2}=m_Z$ and $\alpha_s(m_Z)=0.118$.
The bands give the range for the theoretical prediction obtained from varying the renormalisation scale
from $\mu=m_Z/2$ to $\mu=2 m_Z$.
}
\label{fig_cambridge}
\end{figure}

\begin{figure}[p]
\begin{center}
\includegraphics[bb= 125 460 490 710,width=0.8\textwidth]{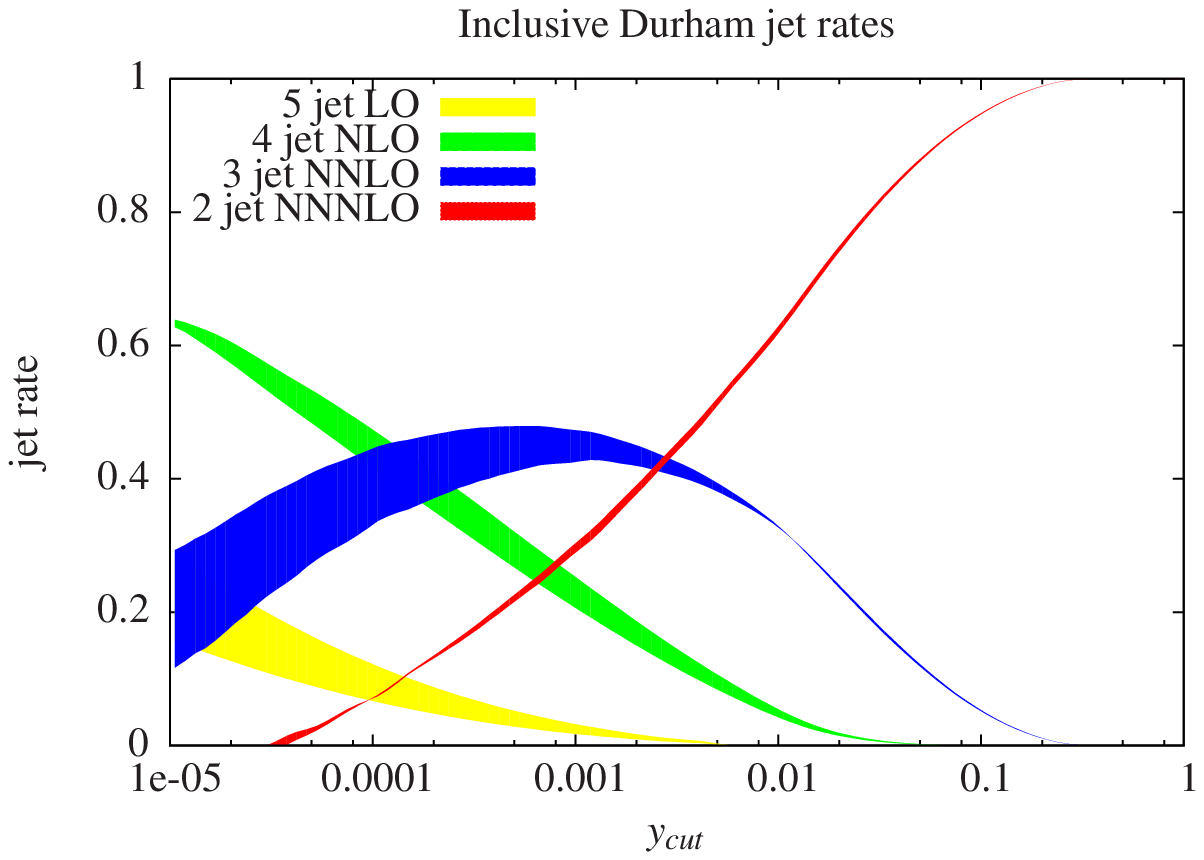}
\includegraphics[bb= 125 460 490 710,width=0.8\textwidth]{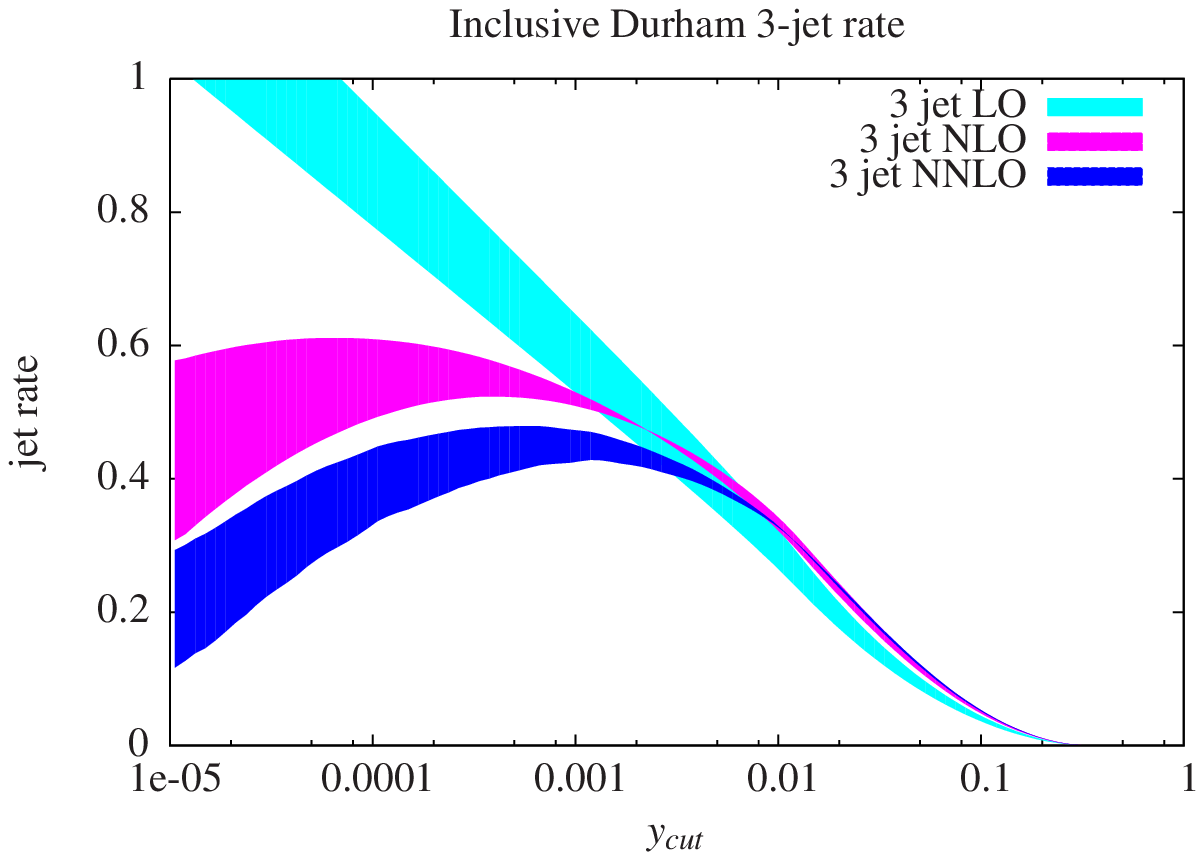}
\end{center}
\caption{
The upper plot shows the jet rates for the inclusive Durham algorithm at order $\alpha_s^3$.
The lower plot shows the three-jet rate for the inclusive Durham algorithm at LO, NLO and NNLO.
All plots are done with $\sqrt{Q^2}=m_Z$ and $\alpha_s(m_Z)=0.118$.
The bands give the range for the theoretical prediction obtained from varying the renormalisation scale
from $\mu=m_Z/2$ to $\mu=2 m_Z$.
}
\label{fig_incldurham}
\end{figure}

\begin{figure}[p]
\begin{center}
\includegraphics[bb= 125 460 490 710,width=0.8\textwidth]{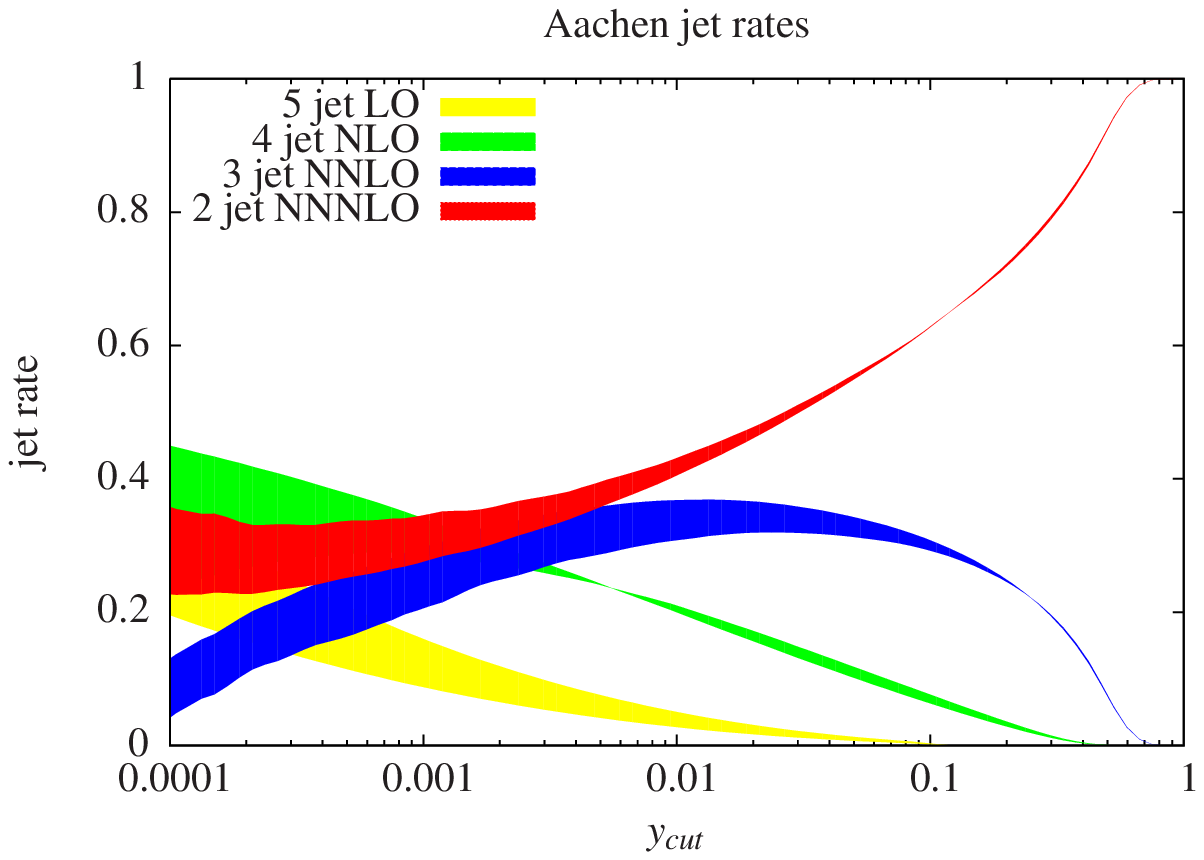}
\includegraphics[bb= 125 460 490 710,width=0.8\textwidth]{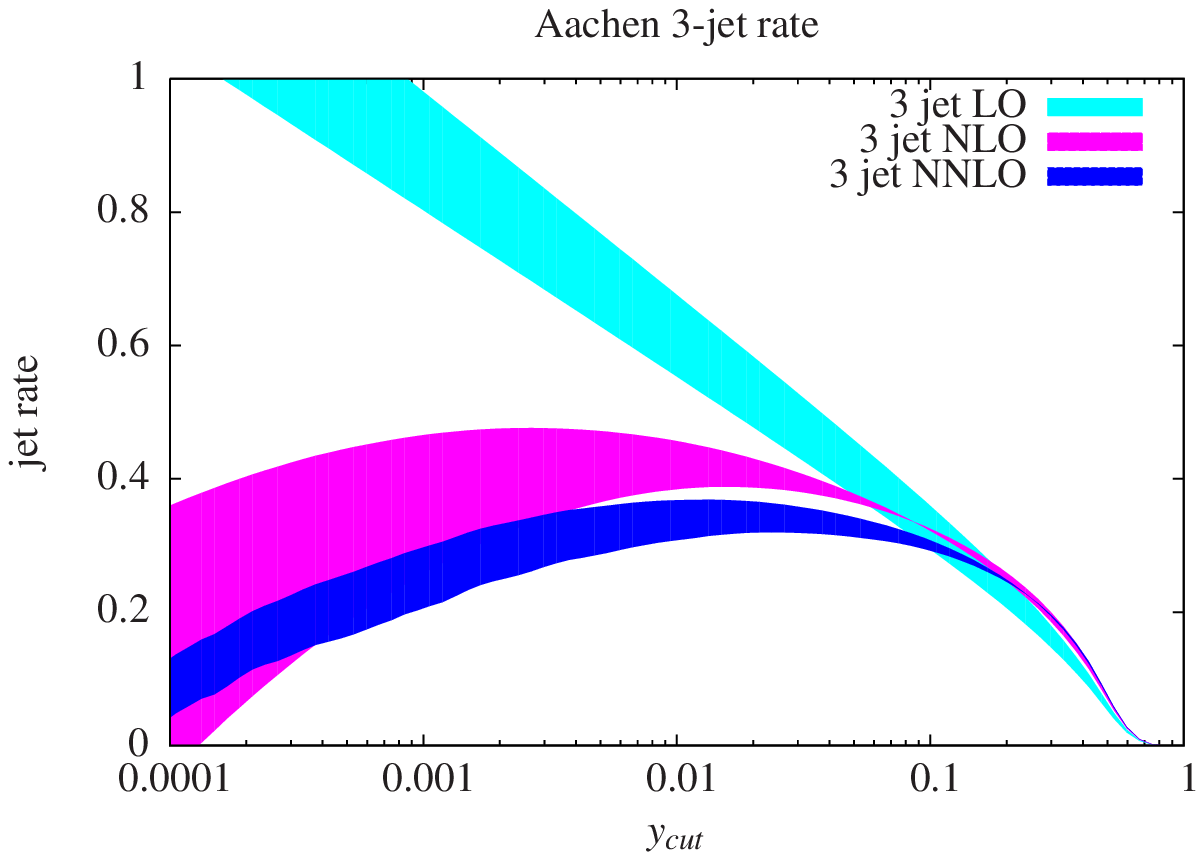}
\end{center}
\caption{
The upper plot shows the jet rates for the Aachen algorithm at order $\alpha_s^3$.
The lower plot shows the three-jet rate for the Aachen algorithm at LO, NLO and NNLO.
All plots are done with $\sqrt{Q^2}=m_Z$ and $\alpha_s(m_Z)=0.118$.
The bands give the range for the theoretical prediction obtained from varying the renormalisation scale
from $\mu=m_Z/2$ to $\mu=2 m_Z$.
}
\label{fig_aachen}
\end{figure}

\begin{figure}[p]
\begin{center}
\includegraphics[bb= 125 460 490 710,width=0.8\textwidth]{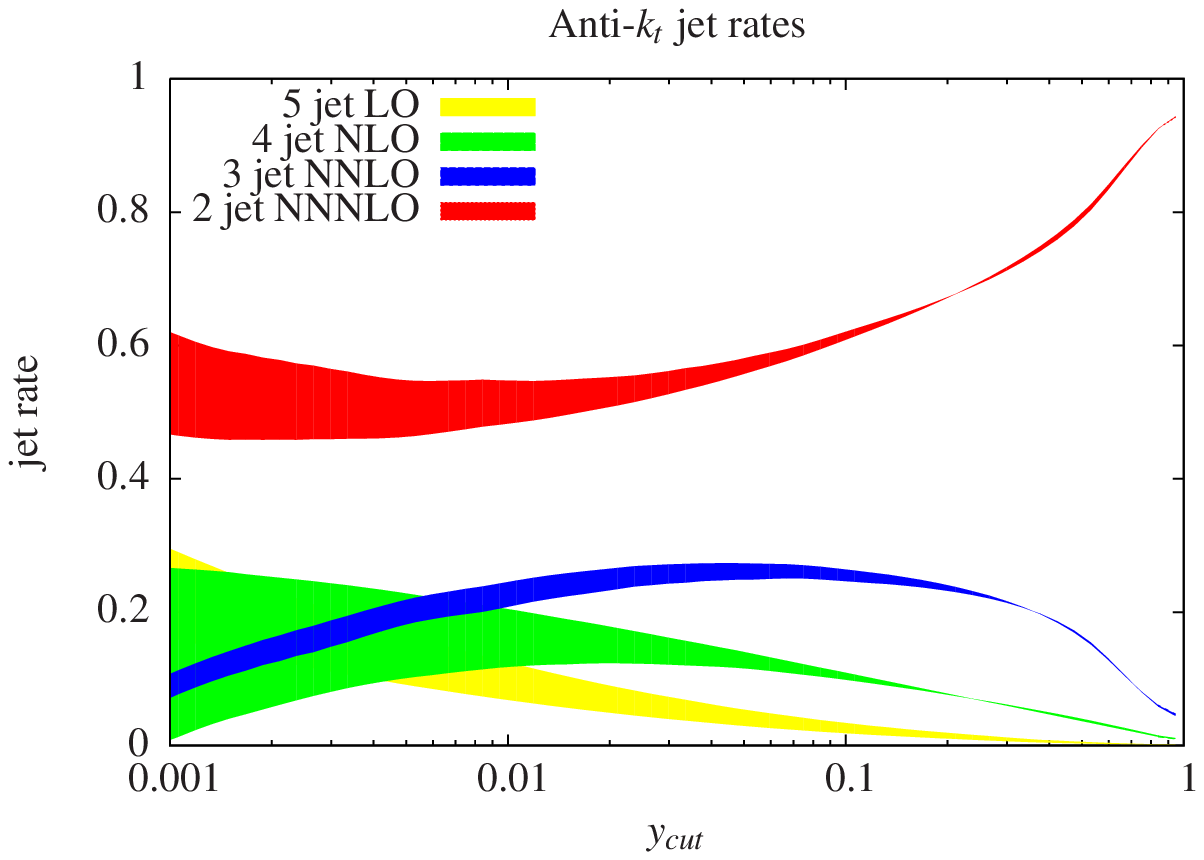}
\includegraphics[bb= 125 460 490 710,width=0.8\textwidth]{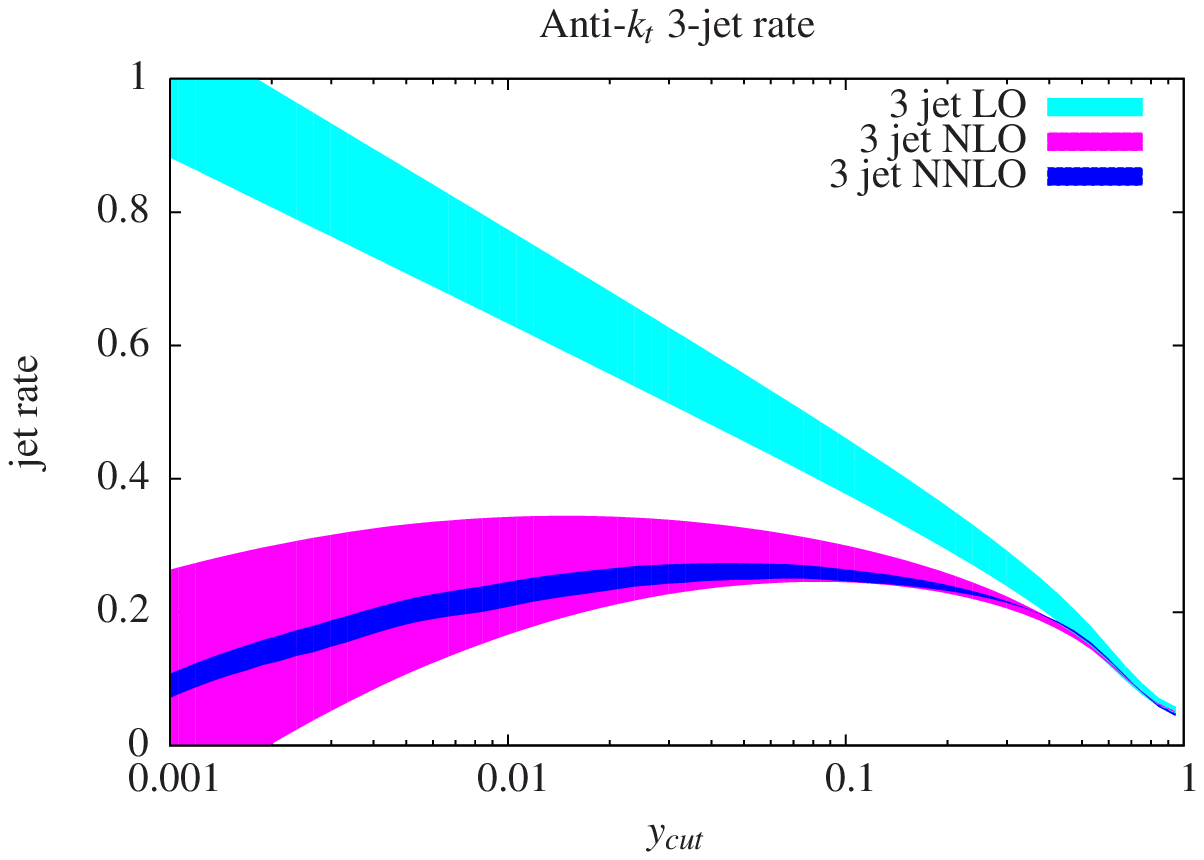}
\end{center}
\caption{
The upper plot shows the jet rates for the anti-$k_t$ algorithm at order $\alpha_s^3$.
The lower plot shows the three-jet rate for the anti-$k_t$ algorithm at LO, NLO and NNLO.
All plots are done with $\sqrt{Q^2}=m_Z$ and $\alpha_s(m_Z)=0.118$.
The bands give the range for the theoretical prediction obtained from varying the renormalisation scale
from $\mu=m_Z/2$ to $\mu=2 m_Z$.
}
\label{fig_antikt}
\end{figure}

\begin{figure}[p]
\begin{center}
\includegraphics[bb= 125 460 490 710,width=0.8\textwidth]{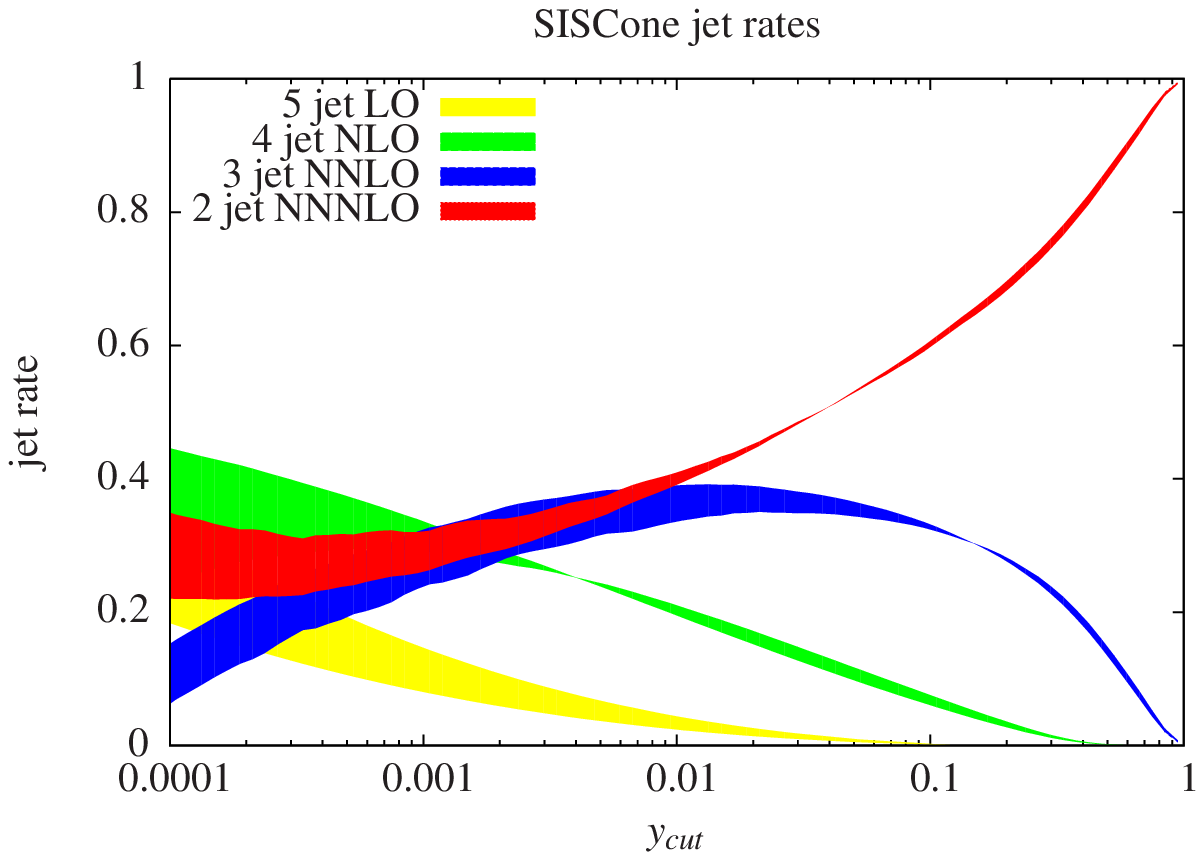}
\includegraphics[bb= 125 460 490 710,width=0.8\textwidth]{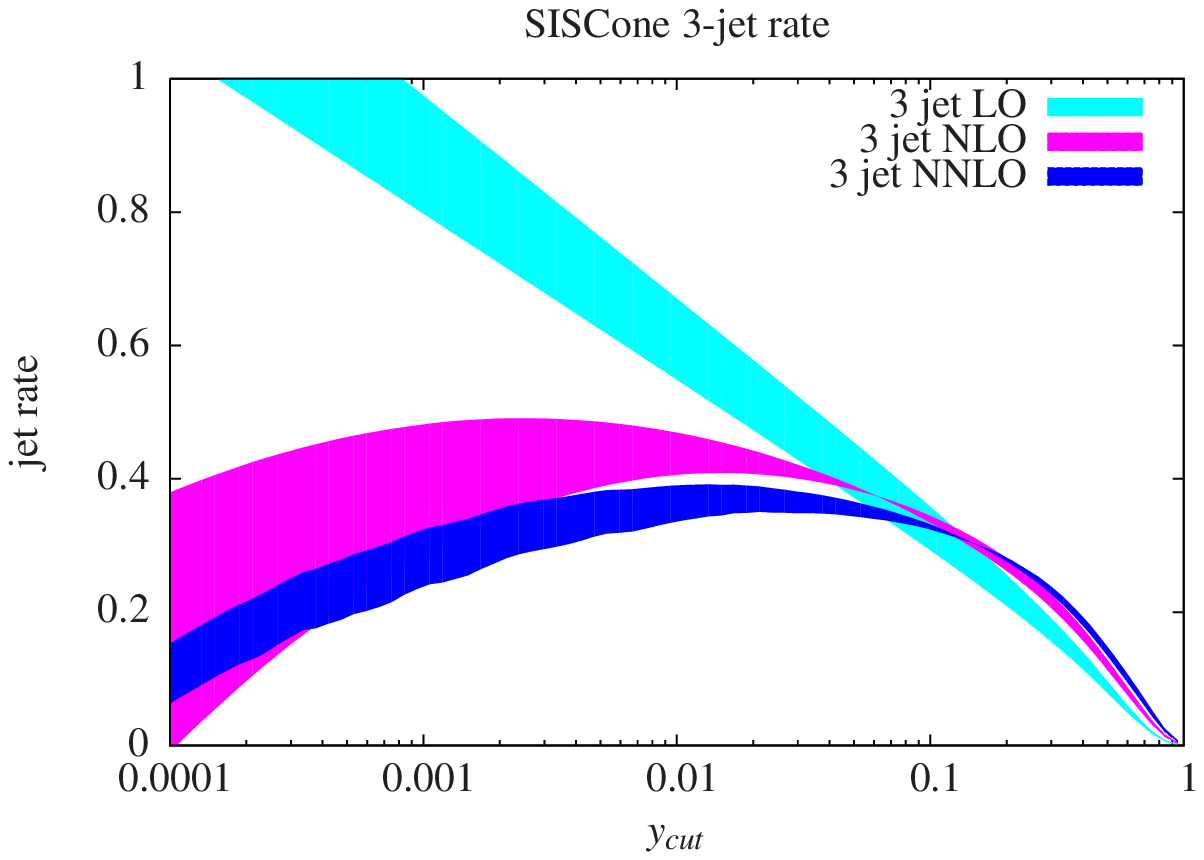}
\end{center}
\caption{
The upper plot shows the jet rates for the SISCone algorithm at order $\alpha_s^3$.
The lower plot shows the three-jet rate for the SISCone algorithm at LO, NLO and NNLO.
All plots are done with $\sqrt{Q^2}=m_Z$ and $\alpha_s(m_Z)=0.118$.
The bands give the range for the theoretical prediction obtained from varying the renormalisation scale
from $\mu=m_Z/2$ to $\mu=2 m_Z$.
}
\label{fig_siscone}
\end{figure}

In this section I present the results for the jet rates.
The jet rates depend on the jet algorithm. 
Results are provided for all jet algorithms introduced in section~(\ref{sec:def}).
These are the exclusive sequential recombination algorithms Durham, Geneva, Jade-E0 and Cambridge, 
the inclusive sequential recombination algorithms Durham, Aachen and anti-$k_t$,
and the infrared-safe cone algorithm SISCone.
The results are LO for the five-jet rates, NLO for the four-jet rates, NNLO for the three-jet rates
and $\mathrm{N}^3$LO for the two-jet rates.
The exclusive sequential recombination algorithms depend on a single parameter $y_{cut}$.
The inclusive sequential recombination algorithms depend in addition on a second parameter $E_{min}$.
The SISCone algorithm depends on four parameters $y_{cut}$ and $E_{min}$ as above and in addition on the two
parameters $n_{pass}$ and $f$.
For all jet algorithm the jet resolution parameter $y_{cut}$ is varied.
The parameter $E_{min}$, which occurs in the inclusive jet algorithms is kept fixed and set to
\bq
 E_{min} & = & \eta_{min} \sqrt{Q^2},
 \;\;\;\;\;\;
 \eta_{min} = 0.077,
\eq
where $\sqrt{Q^2}$ is the centre-of-mass energy.
This value corresponds to $E_{min}=7\;\mbox{GeV}$ for $\sqrt{Q^2}=m_Z$ and is motivated by the value
used by the OPAL collaboration \cite{Akers:1994wj}.
The two additional parameters for the SISCone algorithm are also kept fixed to their default values:
\bq
 n_{pass}=\infty, & & f=0.5.
\eq
For each jet algorithm I show two plots.
The first plot shows the two-, three-, four- and five-jet rate of the jet algorithm at order $\alpha_s^3$
as a function of $y_{cut}$.
The second plot compares the three-jet rate at LO (which is of order $\alpha_s$), at NLO (order $\alpha_s^2$)
and NNLO (order $\alpha_s^3$).
In these plots the values
\bq
 \sqrt{Q^2} = m_Z,
 & & 
 \alpha_s = 0.118
\eq
are used.
The renormalisation scale $\mu$ is varied from $\mu=m_Z/2$ to $\mu=2 m_Z$, this defines a band for the
theoretical prediction which is shown in the plots.
The plots for the various jet algorithms can be found in figs.~(\ref{fig_durham}) - (\ref{fig_siscone}).
It can be seen from the plots that the four inclusive jet algorithms show a different behaviour 
in the small $y_{cut}$ region as compared to the exclusive jet algorithms.
This is expected since the additional (and fixed) parameter $E_{min}$ acts as an additional soft cut-off.

The perturbative coefficients $A_3$, $B_3$, $C_3$, $B_4$, $C_4$ and $C_5$ are reported for each jet algorithm
in tables~(\ref{table1_durham}) - (\ref{table2_siscone}).
Each table contains the results for the $y_{cut}$-values
\bq
 \log_{10} \left( y_{cut} \right) & = & -4.975 + 0.05 i,
 \;\;\;\;\;\; i \in \{0,1,...,99\}.
\eq
Results on the exclusive jet rates have already appeared for a few selected values of $y_{cut}$ 
in \cite{GehrmannDeRidder:2008ug,Weinzierl:2008iv,Weinzierl:2009ms}.
In this article we cover a wider range of $y_{cut}$-values with a finer binning.
The values in this article have been calculated with a significantly higher statistics in the Monte Carlo integration
as compared to the values in \cite{Weinzierl:2008iv,Weinzierl:2009ms}. 
They are in reasonable agreement with the previous values.
The results on the inclusive jet rates are new results.

A few comments are in order: The experimental measured values of the jet rates are by definition
in the range between zero and one.
However, a theoretical prediction based on fixed-order perturbation theory can lead to results which
are negative or larger than one.
This can be seen for example in fig.~(\ref{fig_durham}), showing the perturbative results for the 
exclusive Durham jet rates. In this plot one sees that the leading-order five-jet rate exceeds one for small
values of $y_{cut}$, as well as that the $\mathrm{N}^3$LO two-jet rate is negative for small values
of $y_{cut}$.
In general, unphysical values for the jet rates indicate that large logarithms occur and invalidate
a fixed-order perturbative expansion.
In the small $y_{cut}$-regions a fixed-order calculation has to be combined with a resummation of large logarithms.
In the plots of figs.~(\ref{fig_durham}) - (\ref{fig_siscone}) we restricted our attention 
to the regions of $y_{cut}$, where the fixed-order calculation is expected to be reliable by requiring
that all jet rates are between zero and one.
For each jet algorithm this determines a range $[y_{cut,min},1]$, which is plotted.
Note that the range of $y_{cut}$-values is different for each jet algorithm.
On the other hand we report in tables~(\ref{table1_durham}) - (\ref{table2_siscone})
the perturbative results for all values of $y_{cut}$ between $10^{-5}$ and $1$ for all jet algorithms.
The small $y_{cut}$-values are useful for a matching between fixed-order and resummed calculations,
as well as for a numerical treatment of resummation.

The bands in the plots of figs.~(\ref{fig_durham}) - (\ref{fig_siscone}) are obtained by varying the
the renormalisation scale $\mu$ between $\mu=m_Z/2$ and $\mu=2 m_Z$.
In higher orders of perturbation theory one observes a cross-over of these bands.
As a word of warning I would like to mention, that the usual procedure of estimating the error
of uncalculated higher-order corrections from the scale variation can lead in the cross-over regions
to an under-estimation.
This can be seen in the lower plot of fig.~(\ref{fig_durham}): The NLO prediction for the three-jet rate
shows a cross-over between $y_{cut}=0.001$ and $y_{cut}=0.01$.
However, the NNLO prediction is outside the NLO-band.
A more conservative approach would first construct a hull for the scale-variation bands and estimate
the uncalculated higher-order corrections from this hull.

\section{Conclusions}
\label{sec:conclusions}

In this article I reported on perturbative predictions for the jet rates in electron-positron annihilation.
Eight different jet algorithms have been studied. 
These are the exclusive sequential recombination algorithms Durham, Geneva, Jade-E0 and Cambridge, 
the inclusive sequential recombination algorithms Durham, Aachen and anti-$k_t$,
as well as the infrared-safe cone algorithm SISCone.
The results are obtained in perturbative QCD and are $\mathrm{N}^3$LO for the two-jet rates,
NNLO for the three-jet rates, NLO for the four-jet rates and LO for the five-jet rates.
The results of this paper will be useful for an extraction of $\alpha_s$ from three-jet events in 
electron-positron annihilation.
They are also useful for a study of the properties of inclusive jet algorithms, which is relevant to LHC experiments.

\subsection*{Acknowledgements}

I would like to thank S. Kluth and J.~Schieck
for useful discussions.
The computer support from the Max-Planck-Institut for Physics is greatly acknowledged.

\section*{Erratum}
\label{sec:erratum}

After publication of this paper a bug in the numerical program, which has been used to produce the numerical
results of this paper, has been discovered.
The bug affected  the leading-colour contribution of the $\alpha_s^3$ terms.
A typo in the phase-space parametrisation of the five-parton contribution led to the effect, that
a certain region of phase-space was counted twice, while another region of phase-space was left out.
The bug has not been detected previously, mainly because the wrong phase-space parametrisation reproduced
the correct phase-space volume.
The bug has been found by a re-calculation of the four-jet rates with a new method based on numerical
integration of the virtual loop amplitudes \cite{Assadsolimani:2009cz,Becker:2010ng}.
This bug has now been corrected.
It turns out, that the changes in the results for the eventshapes and the moments of the eventshapes are 
not significant and the corresponding numbers in ref.~\cite{Weinzierl:2009ms,Weinzierl:2009yz}
need not be updated.
However, the changes in the results for the jet rates -- and in particular the changes in the results
for the four-jet rates -- are sizeable.
Therefore, the numerical results and the plots in this version have been corrected.


\bibliography{/home/stefanw/notes/biblio}
\bibliographystyle{/home/stefanw/latex-style/h-physrev3}

%
%

\clearpage

%
%
\begin{table}[p]
\begin{center}
{\scriptsize
\begin{tabular}{|c|c|c|c|c|c|c|}
\hline
 $y_{cut}$ & $A_3$ & $B_3$ & $C_3$ & $B_4$ & $C_4$ & $C_5$ \\
\hline 
$1.05925 \cdot 10^{-5}$ & $1.3562(2) \cdot 10^{2}$ & $-1.4523(6) \cdot 10^{4}$ & $1.44(3) \cdot 10^{5}$ & $1.0683(3) \cdot 10^{4}$ & $-8.46(1) \cdot 10^{5}$ & $6.283(5) \cdot 10^{5}$ \\
$1.18850 \cdot 10^{-5}$ & $1.3258(2) \cdot 10^{2}$ & $-1.3786(6) \cdot 10^{4}$ & $1.17(3) \cdot 10^{5}$ & $1.0188(2) \cdot 10^{4}$ & $-7.73(1) \cdot 10^{5}$ & $5.832(4) \cdot 10^{5}$ \\
$1.33352 \cdot 10^{-5}$ & $1.2958(2) \cdot 10^{2}$ & $-1.3069(5) \cdot 10^{4}$ & $9.5(2) \cdot 10^{4}$ & $9.711(2) \cdot 10^{3}$ & $-7.03(1) \cdot 10^{5}$ & $5.406(4) \cdot 10^{5}$ \\
$1.49624 \cdot 10^{-5}$ & $1.2661(2) \cdot 10^{2}$ & $-1.2379(5) \cdot 10^{4}$ & $7.3(2) \cdot 10^{4}$ & $9.250(2) \cdot 10^{3}$ & $-6.40(1) \cdot 10^{5}$ & $5.007(4) \cdot 10^{5}$ \\
$1.67880 \cdot 10^{-5}$ & $1.2368(2) \cdot 10^{2}$ & $-1.1720(5) \cdot 10^{4}$ & $5.4(2) \cdot 10^{4}$ & $8.806(2) \cdot 10^{3}$ & $-5.81(1) \cdot 10^{5}$ & $4.634(3) \cdot 10^{5}$ \\
$1.88365 \cdot 10^{-5}$ & $1.2080(2) \cdot 10^{2}$ & $-1.1080(5) \cdot 10^{4}$ & $3.5(2) \cdot 10^{4}$ & $8.378(2) \cdot 10^{3}$ & $-5.25(1) \cdot 10^{5}$ & $4.282(3) \cdot 10^{5}$ \\
$2.11349 \cdot 10^{-5}$ & $1.1795(2) \cdot 10^{2}$ & $-1.0468(4) \cdot 10^{4}$ & $2.0(2) \cdot 10^{4}$ & $7.966(2) \cdot 10^{3}$ & $-4.77(1) \cdot 10^{5}$ & $3.953(3) \cdot 10^{5}$ \\
\hline
\end{tabular}
}
\caption{\label{table1_durham}
Perturbative coefficients for the Durham jet rates.
}
\end{center}
\end{table}
\begin{table}[p]
\begin{center}
{\scriptsize
\begin{tabular}{|c|c|c|c|c|c|c|}
\hline
 $y_{cut}$ & $A_3$ & $B_3$ & $C_3$ & $B_4$ & $C_4$ & $C_5$ \\
\hline 
$3.34965 \cdot 10^{-3}$ & $2.6665(1) \cdot 10^{1}$ & $2.29(1) \cdot 10^{1}$ & $-7.3(2) \cdot 10^{3}$ & $3.0325(3) \cdot 10^{2}$ & $8.55(3) \cdot 10^{3}$ & $1.556(4) \cdot 10^{3}$ \\
$3.75837 \cdot 10^{-3}$ & $2.53644(9) \cdot 10^{1}$ & $4.98(1) \cdot 10^{1}$ & $-6.3(2) \cdot 10^{3}$ & $2.6985(3) \cdot 10^{2}$ & $7.86(2) \cdot 10^{3}$ & $1.224(3) \cdot 10^{3}$ \\
$4.21696 \cdot 10^{-3}$ & $2.40981(8) \cdot 10^{1}$ & $7.31(1) \cdot 10^{1}$ & $-5.3(2) \cdot 10^{3}$ & $2.3929(3) \cdot 10^{2}$ & $7.20(2) \cdot 10^{3}$ & $9.62(3) \cdot 10^{2}$ \\
$4.73151 \cdot 10^{-3}$ & $2.28662(7) \cdot 10^{1}$ & $9.29(1) \cdot 10^{1}$ & $-4.3(2) \cdot 10^{3}$ & $2.1141(2) \cdot 10^{2}$ & $6.54(2) \cdot 10^{3}$ & $7.67(2) \cdot 10^{2}$ \\
$5.30884 \cdot 10^{-3}$ & $2.16676(7) \cdot 10^{1}$ & $1.094(1) \cdot 10^{2}$ & $-3.5(1) \cdot 10^{3}$ & $1.8600(2) \cdot 10^{2}$ & $5.89(2) \cdot 10^{3}$ & $5.97(2) \cdot 10^{2}$ \\
$5.95662 \cdot 10^{-3}$ & $2.05019(6) \cdot 10^{1}$ & $1.228(1) \cdot 10^{2}$ & $-2.8(1) \cdot 10^{3}$ & $1.6296(2) \cdot 10^{2}$ & $5.30(2) \cdot 10^{3}$ & $4.48(1) \cdot 10^{2}$ \\
$6.68344 \cdot 10^{-3}$ & $1.93703(6) \cdot 10^{1}$ & $1.3349(9) \cdot 10^{2}$ & $-2.1(1) \cdot 10^{3}$ & $1.4209(2) \cdot 10^{2}$ & $4.69(1) \cdot 10^{3}$ & $3.45(1) \cdot 10^{2}$ \\
\hline
\end{tabular}
}
\caption{\label{table2_durham}
Perturbative coefficients for the Durham jet rates (continued).
}
\end{center}
\end{table}
\clearpage
\begin{table}[p]
\begin{center}
{\scriptsize
\begin{tabular}{|c|c|c|c|c|c|c|}
\hline
 $y_{cut}$ & $A_3$ & $B_3$ & $C_3$ & $B_4$ & $C_4$ & $C_5$ \\
\hline 
$1.05925 \cdot 10^{-5}$ & $3.366(1) \cdot 10^{2}$ & $-1.241(1) \cdot 10^{5}$ & $1.661(5) \cdot 10^{7}$ & $8.486(7) \cdot 10^{4}$ & $-3.395(4) \cdot 10^{7}$ & $1.924(2) \cdot 10^{7}$ \\
$1.18850 \cdot 10^{-5}$ & $3.297(1) \cdot 10^{2}$ & $-1.191(1) \cdot 10^{5}$ & $1.550(4) \cdot 10^{7}$ & $8.174(7) \cdot 10^{4}$ & $-3.200(4) \cdot 10^{7}$ & $1.826(2) \cdot 10^{7}$ \\
$1.33352 \cdot 10^{-5}$ & $3.228(1) \cdot 10^{2}$ & $-1.1435(9) \cdot 10^{5}$ & $1.445(4) \cdot 10^{7}$ & $7.869(6) \cdot 10^{4}$ & $-3.016(4) \cdot 10^{7}$ & $1.729(2) \cdot 10^{7}$ \\
$1.49624 \cdot 10^{-5}$ & $3.161(1) \cdot 10^{2}$ & $-1.0965(9) \cdot 10^{5}$ & $1.347(4) \cdot 10^{7}$ & $7.565(6) \cdot 10^{4}$ & $-2.846(4) \cdot 10^{7}$ & $1.637(2) \cdot 10^{7}$ \\
$1.67880 \cdot 10^{-5}$ & $3.094(1) \cdot 10^{2}$ & $-1.0507(9) \cdot 10^{5}$ & $1.256(4) \cdot 10^{7}$ & $7.271(6) \cdot 10^{4}$ & $-2.676(3) \cdot 10^{7}$ & $1.547(1) \cdot 10^{7}$ \\
$1.88365 \cdot 10^{-5}$ & $3.027(1) \cdot 10^{2}$ & $-1.0070(8) \cdot 10^{5}$ & $1.167(3) \cdot 10^{7}$ & $6.986(6) \cdot 10^{4}$ & $-2.513(3) \cdot 10^{7}$ & $1.459(1) \cdot 10^{7}$ \\
$2.11349 \cdot 10^{-5}$ & $2.962(1) \cdot 10^{2}$ & $-9.641(8) \cdot 10^{4}$ & $1.086(3) \cdot 10^{7}$ & $6.704(5) \cdot 10^{4}$ & $-2.354(3) \cdot 10^{7}$ & $1.375(1) \cdot 10^{7}$ \\
\hline
\end{tabular}
}
\caption{\label{table1_geneva}
Perturbative coefficients for the Geneva jet rates.
}
\end{center}
\end{table}
\begin{table}[p]
\begin{center}
{\scriptsize
\begin{tabular}{|c|c|c|c|c|c|c|}
\hline
 $y_{cut}$ & $A_3$ & $B_3$ & $C_3$ & $B_4$ & $C_4$ & $C_5$ \\
\hline 
$3.34965 \cdot 10^{-3}$ & $7.854(2) \cdot 10^{1}$ & $-5.290(4) \cdot 10^{3}$ & $-2.2(1) \cdot 10^{4}$ & $4.555(1) \cdot 10^{3}$ & $-2.328(6) \cdot 10^{5}$ & $2.197(2) \cdot 10^{5}$ \\
$3.75837 \cdot 10^{-3}$ & $7.518(2) \cdot 10^{1}$ & $-4.764(3) \cdot 10^{3}$ & $-2.9(1) \cdot 10^{4}$ & $4.158(1) \cdot 10^{3}$ & $-1.944(5) \cdot 10^{5}$ & $1.902(2) \cdot 10^{5}$ \\
$4.21696 \cdot 10^{-3}$ & $7.189(2) \cdot 10^{1}$ & $-4.287(3) \cdot 10^{3}$ & $-3.4(1) \cdot 10^{4}$ & $3.7844(9) \cdot 10^{3}$ & $-1.611(4) \cdot 10^{5}$ & $1.641(1) \cdot 10^{5}$ \\
$4.73151 \cdot 10^{-3}$ & $6.869(2) \cdot 10^{1}$ & $-3.840(3) \cdot 10^{3}$ & $-3.54(9) \cdot 10^{4}$ & $3.4362(8) \cdot 10^{3}$ & $-1.325(4) \cdot 10^{5}$ & $1.410(1) \cdot 10^{5}$ \\
$5.30884 \cdot 10^{-3}$ & $6.555(1) \cdot 10^{1}$ & $-3.430(2) \cdot 10^{3}$ & $-3.74(9) \cdot 10^{4}$ & $3.1103(7) \cdot 10^{3}$ & $-1.076(3) \cdot 10^{5}$ & $1.207(1) \cdot 10^{5}$ \\
$5.95662 \cdot 10^{-3}$ & $6.248(1) \cdot 10^{1}$ & $-3.053(2) \cdot 10^{3}$ & $-3.88(8) \cdot 10^{4}$ & $2.8084(6) \cdot 10^{3}$ & $-8.60(3) \cdot 10^{4}$ & $1.0255(9) \cdot 10^{5}$ \\
$6.68344 \cdot 10^{-3}$ & $5.948(1) \cdot 10^{1}$ & $-2.706(2) \cdot 10^{3}$ & $-3.80(7) \cdot 10^{4}$ & $2.5288(5) \cdot 10^{3}$ & $-6.85(3) \cdot 10^{4}$ & $8.687(8) \cdot 10^{4}$ \\
\hline
\end{tabular}
}
\caption{\label{table2_geneva}
Perturbative coefficients for the Geneva jet rates (continued).
}
\end{center}
\end{table}
\clearpage
\begin{table}[p]
\begin{center}
{\scriptsize
\begin{tabular}{|c|c|c|c|c|c|c|}
\hline
 $y_{cut}$ & $A_3$ & $B_3$ & $C_3$ & $B_4$ & $C_4$ & $C_5$ \\
\hline 
$1.05925 \cdot 10^{-5}$ & $3.028(1) \cdot 10^{2}$ & $-6.834(7) \cdot 10^{4}$ & $5.25(3) \cdot 10^{6}$ & $3.508(2) \cdot 10^{4}$ & $-6.48(1) \cdot 10^{6}$ & $2.936(3) \cdot 10^{6}$ \\
$1.18850 \cdot 10^{-5}$ & $2.962(1) \cdot 10^{2}$ & $-6.519(6) \cdot 10^{4}$ & $4.84(2) \cdot 10^{6}$ & $3.352(2) \cdot 10^{4}$ & $-5.99(1) \cdot 10^{6}$ & $2.733(3) \cdot 10^{6}$ \\
$1.33352 \cdot 10^{-5}$ & $2.898(1) \cdot 10^{2}$ & $-6.213(6) \cdot 10^{4}$ & $4.42(2) \cdot 10^{6}$ & $3.201(2) \cdot 10^{4}$ & $-5.524(9) \cdot 10^{6}$ & $2.541(2) \cdot 10^{6}$ \\
$1.49624 \cdot 10^{-5}$ & $2.833(1) \cdot 10^{2}$ & $-5.916(6) \cdot 10^{4}$ & $4.02(2) \cdot 10^{6}$ & $3.056(1) \cdot 10^{4}$ & $-5.096(8) \cdot 10^{6}$ & $2.362(2) \cdot 10^{6}$ \\
$1.67880 \cdot 10^{-5}$ & $2.7701(9) \cdot 10^{2}$ & $-5.630(6) \cdot 10^{4}$ & $3.66(2) \cdot 10^{6}$ & $2.915(1) \cdot 10^{4}$ & $-4.699(8) \cdot 10^{6}$ & $2.194(2) \cdot 10^{6}$ \\
$1.88365 \cdot 10^{-5}$ & $2.7077(9) \cdot 10^{2}$ & $-5.352(5) \cdot 10^{4}$ & $3.35(2) \cdot 10^{6}$ & $2.779(1) \cdot 10^{4}$ & $-4.322(7) \cdot 10^{6}$ & $2.034(2) \cdot 10^{6}$ \\
$2.11349 \cdot 10^{-5}$ & $2.6459(9) \cdot 10^{2}$ & $-5.083(5) \cdot 10^{4}$ & $3.04(2) \cdot 10^{6}$ & $2.648(1) \cdot 10^{4}$ & $-3.968(7) \cdot 10^{6}$ & $1.886(2) \cdot 10^{6}$ \\
\hline
\end{tabular}
}
\caption{\label{table1_jadeE0}
Perturbative coefficients for the Jade-E0 jet rates.
}
\end{center}
\end{table}
\begin{table}[p]
\begin{center}
{\scriptsize
\begin{tabular}{|c|c|c|c|c|c|c|}
\hline
 $y_{cut}$ & $A_3$ & $B_3$ & $C_3$ & $B_4$ & $C_4$ & $C_5$ \\
\hline 
$3.34965 \cdot 10^{-3}$ & $6.266(1) \cdot 10^{1}$ & $-1.170(3) \cdot 10^{3}$ & $-5.21(8) \cdot 10^{4}$ & $1.1511(2) \cdot 10^{3}$ & $2.43(1) \cdot 10^{4}$ & $1.076(2) \cdot 10^{4}$ \\
$3.75837 \cdot 10^{-3}$ & $5.968(1) \cdot 10^{1}$ & $-9.72(2) \cdot 10^{2}$ & $-4.68(8) \cdot 10^{4}$ & $1.0291(1) \cdot 10^{3}$ & $2.40(1) \cdot 10^{4}$ & $8.80(1) \cdot 10^{3}$ \\
$4.21696 \cdot 10^{-3}$ & $5.6783(9) \cdot 10^{1}$ & $-7.89(2) \cdot 10^{2}$ & $-4.24(7) \cdot 10^{4}$ & $9.168(1) \cdot 10^{2}$ & $2.317(9) \cdot 10^{4}$ & $7.17(1) \cdot 10^{3}$ \\
$4.73151 \cdot 10^{-3}$ & $5.3966(7) \cdot 10^{1}$ & $-6.31(2) \cdot 10^{2}$ & $-3.83(7) \cdot 10^{4}$ & $8.139(1) \cdot 10^{2}$ & $2.222(8) \cdot 10^{4}$ & $5.74(1) \cdot 10^{3}$ \\
$5.30884 \cdot 10^{-3}$ & $5.1221(5) \cdot 10^{1}$ & $-4.82(1) \cdot 10^{2}$ & $-3.30(6) \cdot 10^{4}$ & $7.1973(9) \cdot 10^{2}$ & $2.098(7) \cdot 10^{4}$ & $4.585(8) \cdot 10^{3}$ \\
$5.95662 \cdot 10^{-3}$ & $4.8541(4) \cdot 10^{1}$ & $-3.47(1) \cdot 10^{2}$ & $-2.86(6) \cdot 10^{4}$ & $6.3387(8) \cdot 10^{2}$ & $1.974(6) \cdot 10^{4}$ & $3.632(7) \cdot 10^{3}$ \\
$6.68344 \cdot 10^{-3}$ & $4.5920(3) \cdot 10^{1}$ & $-2.28(1) \cdot 10^{2}$ & $-2.49(5) \cdot 10^{4}$ & $5.5576(7) \cdot 10^{2}$ & $1.827(5) \cdot 10^{4}$ & $2.843(6) \cdot 10^{3}$ \\
\hline
\end{tabular}
}
\caption{\label{table2_jadeE0}
Perturbative coefficients for the Jade-E0 jet rates (continued).
}
\end{center}
\end{table}
\clearpage
\begin{table}[p]
\begin{center}
{\scriptsize
\begin{tabular}{|c|c|c|c|c|c|c|}
\hline
 $y_{cut}$ & $A_3$ & $B_3$ & $C_3$ & $B_4$ & $C_4$ & $C_5$ \\
\hline 
$1.05925 \cdot 10^{-5}$ & $1.3562(2) \cdot 10^{2}$ & $-1.4971(6) \cdot 10^{4}$ & $2.10(3) \cdot 10^{5}$ & $1.0810(3) \cdot 10^{4}$ & $-9.25(1) \cdot 10^{5}$ & $6.473(5) \cdot 10^{5}$ \\
$1.18850 \cdot 10^{-5}$ & $1.3258(2) \cdot 10^{2}$ & $-1.4228(6) \cdot 10^{4}$ & $1.80(3) \cdot 10^{5}$ & $1.0312(2) \cdot 10^{4}$ & $-8.47(1) \cdot 10^{5}$ & $6.009(5) \cdot 10^{5}$ \\
$1.33352 \cdot 10^{-5}$ & $1.2958(2) \cdot 10^{2}$ & $-1.3501(6) \cdot 10^{4}$ & $1.51(2) \cdot 10^{5}$ & $9.830(2) \cdot 10^{3}$ & $-7.74(1) \cdot 10^{5}$ & $5.570(4) \cdot 10^{5}$ \\
$1.49624 \cdot 10^{-5}$ & $1.2661(2) \cdot 10^{2}$ & $-1.2802(5) \cdot 10^{4}$ & $1.22(2) \cdot 10^{5}$ & $9.365(2) \cdot 10^{3}$ & $-7.06(1) \cdot 10^{5}$ & $5.162(4) \cdot 10^{5}$ \\
$1.67880 \cdot 10^{-5}$ & $1.2368(2) \cdot 10^{2}$ & $-1.2132(5) \cdot 10^{4}$ & $1.00(2) \cdot 10^{5}$ & $8.917(2) \cdot 10^{3}$ & $-6.44(1) \cdot 10^{5}$ & $4.778(4) \cdot 10^{5}$ \\
$1.88365 \cdot 10^{-5}$ & $1.2080(2) \cdot 10^{2}$ & $-1.1482(5) \cdot 10^{4}$ & $8.0(2) \cdot 10^{4}$ & $8.486(2) \cdot 10^{3}$ & $-5.84(1) \cdot 10^{5}$ & $4.417(3) \cdot 10^{5}$ \\
$2.11349 \cdot 10^{-5}$ & $1.1795(2) \cdot 10^{2}$ & $-1.0859(4) \cdot 10^{4}$ & $6.3(2) \cdot 10^{4}$ & $8.070(2) \cdot 10^{3}$ & $-5.31(1) \cdot 10^{5}$ & $4.081(3) \cdot 10^{5}$ \\
\hline
\end{tabular}
}
\caption{\label{table1_cambridge}
Perturbative coefficients for the Cambridge jet rates.
}
\end{center}
\end{table}
\begin{table}[p]
\begin{center}
{\scriptsize
\begin{tabular}{|c|c|c|c|c|c|c|}
\hline
 $y_{cut}$ & $A_3$ & $B_3$ & $C_3$ & $B_4$ & $C_4$ & $C_5$ \\
\hline 
$3.34965 \cdot 10^{-3}$ & $2.6665(1) \cdot 10^{1}$ & $-6.26(2) \cdot 10^{1}$ & $-8.3(2) \cdot 10^{3}$ & $3.1235(3) \cdot 10^{2}$ & $6.92(3) \cdot 10^{3}$ & $1.647(4) \cdot 10^{3}$ \\
$3.75837 \cdot 10^{-3}$ & $2.53644(9) \cdot 10^{1}$ & $-3.11(1) \cdot 10^{1}$ & $-7.6(2) \cdot 10^{3}$ & $2.7810(3) \cdot 10^{2}$ & $6.49(2) \cdot 10^{3}$ & $1.295(3) \cdot 10^{3}$ \\
$4.21696 \cdot 10^{-3}$ & $2.40981(8) \cdot 10^{1}$ & $-3.6(1) \cdot 10^{0}$ & $-6.7(2) \cdot 10^{3}$ & $2.4676(3) \cdot 10^{2}$ & $5.97(2) \cdot 10^{3}$ & $1.045(3) \cdot 10^{3}$ \\
$4.73151 \cdot 10^{-3}$ & $2.28662(7) \cdot 10^{1}$ & $2.04(1) \cdot 10^{1}$ & $-5.7(2) \cdot 10^{3}$ & $2.1812(2) \cdot 10^{2}$ & $5.48(2) \cdot 10^{3}$ & $8.37(2) \cdot 10^{2}$ \\
$5.30884 \cdot 10^{-3}$ & $2.16676(7) \cdot 10^{1}$ & $4.10(1) \cdot 10^{1}$ & $-4.9(1) \cdot 10^{3}$ & $1.9202(2) \cdot 10^{2}$ & $4.96(2) \cdot 10^{3}$ & $6.51(2) \cdot 10^{2}$ \\
$5.95662 \cdot 10^{-3}$ & $2.05019(6) \cdot 10^{1}$ & $5.84(1) \cdot 10^{1}$ & $-4.2(1) \cdot 10^{3}$ & $1.6832(2) \cdot 10^{2}$ & $4.56(2) \cdot 10^{3}$ & $5.14(2) \cdot 10^{2}$ \\
$6.68344 \cdot 10^{-3}$ & $1.93703(6) \cdot 10^{1}$ & $7.296(9) \cdot 10^{1}$ & $-3.5(1) \cdot 10^{3}$ & $1.4685(2) \cdot 10^{2}$ & $4.06(1) \cdot 10^{3}$ & $3.93(1) \cdot 10^{2}$ \\
\hline
\end{tabular}
}
\caption{\label{table2_cambridge}
Perturbative coefficients for the Cambridge jet rates (continued).
}
\end{center}
\end{table}
\clearpage
\begin{table}[p]
\begin{center}
{\scriptsize
\begin{tabular}{|c|c|c|c|c|c|c|}
\hline
 $y_{cut}$ & $A_3$ & $B_3$ & $C_3$ & $B_4$ & $C_4$ & $C_5$ \\
\hline 
$1.05925 \cdot 10^{-5}$ & $5.9831(5) \cdot 10^{1}$ & $-1.725(2) \cdot 10^{3}$ & $-4.5(1) \cdot 10^{4}$ & $1.8179(4) \cdot 10^{3}$ & $5.7(3) \cdot 10^{3}$ & $3.156(5) \cdot 10^{4}$ \\
$1.18850 \cdot 10^{-5}$ & $5.9103(5) \cdot 10^{1}$ & $-1.664(2) \cdot 10^{3}$ & $-4.3(1) \cdot 10^{4}$ & $1.7725(4) \cdot 10^{3}$ & $6.6(3) \cdot 10^{3}$ & $3.039(5) \cdot 10^{4}$ \\
$1.33352 \cdot 10^{-5}$ & $5.8375(4) \cdot 10^{1}$ & $-1.605(2) \cdot 10^{3}$ & $-4.3(1) \cdot 10^{4}$ & $1.7279(4) \cdot 10^{3}$ & $7.5(3) \cdot 10^{3}$ & $2.924(5) \cdot 10^{4}$ \\
$1.49624 \cdot 10^{-5}$ & $5.7649(4) \cdot 10^{1}$ & $-1.546(2) \cdot 10^{3}$ & $-4.3(1) \cdot 10^{4}$ & $1.6838(4) \cdot 10^{3}$ & $8.6(3) \cdot 10^{3}$ & $2.812(5) \cdot 10^{4}$ \\
$1.67880 \cdot 10^{-5}$ & $5.6921(4) \cdot 10^{1}$ & $-1.491(2) \cdot 10^{3}$ & $-4.3(1) \cdot 10^{4}$ & $1.6401(4) \cdot 10^{3}$ & $9.4(3) \cdot 10^{3}$ & $2.706(4) \cdot 10^{4}$ \\
$1.88365 \cdot 10^{-5}$ & $5.6192(4) \cdot 10^{1}$ & $-1.433(2) \cdot 10^{3}$ & $-4.1(1) \cdot 10^{4}$ & $1.5972(4) \cdot 10^{3}$ & $1.01(2) \cdot 10^{4}$ & $2.601(4) \cdot 10^{4}$ \\
$2.11349 \cdot 10^{-5}$ & $5.5465(4) \cdot 10^{1}$ & $-1.377(1) \cdot 10^{3}$ & $-4.05(9) \cdot 10^{4}$ & $1.5548(4) \cdot 10^{3}$ & $1.06(2) \cdot 10^{4}$ & $2.498(4) \cdot 10^{4}$ \\
\hline
\end{tabular}
}
\caption{\label{table1_incldurham}
Perturbative coefficients for the inclusive Durham jet rates.
}
\end{center}
\end{table}
\begin{table}[p]
\begin{center}
{\scriptsize
\begin{tabular}{|c|c|c|c|c|c|c|}
\hline
 $y_{cut}$ & $A_3$ & $B_3$ & $C_3$ & $B_4$ & $C_4$ & $C_5$ \\
\hline 
$3.34965 \cdot 10^{-3}$ & $2.31473(9) \cdot 10^{1}$ & $8.20(1) \cdot 10^{1}$ & $-4.6(2) \cdot 10^{3}$ & $2.3192(3) \cdot 10^{2}$ & $7.01(2) \cdot 10^{3}$ & $9.93(3) \cdot 10^{2}$ \\
$3.75837 \cdot 10^{-3}$ & $2.23800(8) \cdot 10^{1}$ & $9.32(1) \cdot 10^{1}$ & $-4.1(2) \cdot 10^{3}$ & $2.1395(2) \cdot 10^{2}$ & $6.58(2) \cdot 10^{3}$ & $8.41(2) \cdot 10^{2}$ \\
$4.21696 \cdot 10^{-3}$ & $2.16073(8) \cdot 10^{1}$ & $1.037(1) \cdot 10^{2}$ & $-3.6(1) \cdot 10^{3}$ & $1.9651(2) \cdot 10^{2}$ & $6.15(2) \cdot 10^{3}$ & $7.18(2) \cdot 10^{2}$ \\
$4.73151 \cdot 10^{-3}$ & $2.08290(7) \cdot 10^{1}$ & $1.133(1) \cdot 10^{2}$ & $-3.0(1) \cdot 10^{3}$ & $1.7962(2) \cdot 10^{2}$ & $5.72(2) \cdot 10^{3}$ & $6.05(2) \cdot 10^{2}$ \\
$5.30884 \cdot 10^{-3}$ & $2.00428(6) \cdot 10^{1}$ & $1.220(1) \cdot 10^{2}$ & $-2.7(1) \cdot 10^{3}$ & $1.6325(2) \cdot 10^{2}$ & $5.28(2) \cdot 10^{3}$ & $6.551(2) \cdot 10^{2}$ \\
$5.95662 \cdot 10^{-3}$ & $1.92479(6) \cdot 10^{1}$ & $1.2974(9) \cdot 10^{2}$ & $-2.2(1) \cdot 10^{3}$ & $1.4743(2) \cdot 10^{2}$ & $4.86(1) \cdot 10^{3}$ & $5.408(2) \cdot 10^{2}$ \\
$6.68344 \cdot 10^{-3}$ & $1.84436(5) \cdot 10^{1}$ & $1.3639(9) \cdot 10^{2}$ & $-1.7(1) \cdot 10^{3}$ & $1.3217(2) \cdot 10^{2}$ & $4.40(1) \cdot 10^{3}$ & $4.387(1) \cdot 10^{2}$ \\
\hline
\end{tabular}
}
\caption{\label{table2_incldurham}
Perturbative coefficients for the inclusive Durham jet rates (continued).
}
\end{center}
\end{table}
\clearpage
\begin{table}[p]
\begin{center}
{\scriptsize
\begin{tabular}{|c|c|c|c|c|c|c|}
\hline
 $y_{cut}$ & $A_3$ & $B_3$ & $C_3$ & $B_4$ & $C_4$ & $C_5$ \\
\hline 
$1.05925 \cdot 10^{-5}$ & $7.5753(6) \cdot 10^{1}$ & $-4.143(5) \cdot 10^{3}$ & $-1.9(3) \cdot 10^{4}$ & $3.0749(7) \cdot 10^{3}$ & $-1.028(8) \cdot 10^{5}$ & $7.20(1) \cdot 10^{4}$ \\
$1.18850 \cdot 10^{-5}$ & $7.5026(5) \cdot 10^{1}$ & $-4.059(5) \cdot 10^{3}$ & $-1.8(3) \cdot 10^{4}$ & $3.0162(7) \cdot 10^{3}$ & $-9.95(8) \cdot 10^{4}$ & $7.01(1) \cdot 10^{4}$ \\
$1.33352 \cdot 10^{-5}$ & $7.4299(5) \cdot 10^{1}$ & $-3.976(5) \cdot 10^{3}$ & $-1.9(3) \cdot 10^{4}$ & $2.9580(7) \cdot 10^{3}$ & $-9.62(8) \cdot 10^{4}$ & $6.81(1) \cdot 10^{4}$ \\
$1.49624 \cdot 10^{-5}$ & $7.3571(5) \cdot 10^{1}$ & $-3.891(5) \cdot 10^{3}$ & $-1.9(3) \cdot 10^{4}$ & $2.9004(7) \cdot 10^{3}$ & $-9.32(7) \cdot 10^{4}$ & $6.62(1) \cdot 10^{4}$ \\
$1.67880 \cdot 10^{-5}$ & $7.2843(5) \cdot 10^{1}$ & $-3.806(5) \cdot 10^{3}$ & $-2.0(3) \cdot 10^{4}$ & $2.8432(7) \cdot 10^{3}$ & $-8.98(7) \cdot 10^{4}$ & $6.44(1) \cdot 10^{4}$ \\
$1.88365 \cdot 10^{-5}$ & $7.2116(5) \cdot 10^{1}$ & $-3.722(4) \cdot 10^{3}$ & $-2.2(3) \cdot 10^{4}$ & $2.7868(7) \cdot 10^{3}$ & $-8.70(7) \cdot 10^{4}$ & $6.26(1) \cdot 10^{4}$ \\
$2.11349 \cdot 10^{-5}$ & $7.1388(5) \cdot 10^{1}$ & $-3.640(4) \cdot 10^{3}$ & $-2.3(3) \cdot 10^{4}$ & $2.7305(7) \cdot 10^{3}$ & $-8.35(7) \cdot 10^{4}$ & $6.07(1) \cdot 10^{4}$ \\
\hline
\end{tabular}
}
\caption{\label{table1_aachen}
Perturbative coefficients for the Aachen jet rates.
}
\end{center}
\end{table}
\begin{table}[p]
\begin{center}
{\scriptsize
\begin{tabular}{|c|c|c|c|c|c|c|}
\hline
 $y_{cut}$ & $A_3$ & $B_3$ & $C_3$ & $B_4$ & $C_4$ & $C_5$ \\
\hline 
$3.34965 \cdot 10^{-3}$ & $3.9361(3) \cdot 10^{1}$ & $-8.006(6) \cdot 10^{2}$ & $-1.93(4) \cdot 10^{4}$ & $8.233(2) \cdot 10^{2}$ & $-1.22(9) \cdot 10^{3}$ & $1.023(2) \cdot 10^{4}$ \\
$3.75837 \cdot 10^{-3}$ & $3.8632(3) \cdot 10^{1}$ & $-7.572(6) \cdot 10^{2}$ & $-1.91(4) \cdot 10^{4}$ & $7.924(1) \cdot 10^{2}$ & $-5.9(9) \cdot 10^{2}$ & $9.64(2) \cdot 10^{3}$ \\
$4.21696 \cdot 10^{-3}$ & $3.7902(3) \cdot 10^{1}$ & $-7.166(6) \cdot 10^{2}$ & $-1.83(4) \cdot 10^{4}$ & $7.623(1) \cdot 10^{2}$ & $1.7(8) \cdot 10^{2}$ & $9.09(1) \cdot 10^{3}$ \\
$4.73151 \cdot 10^{-3}$ & $3.7170(3) \cdot 10^{1}$ & $-6.756(6) \cdot 10^{2}$ & $-1.83(4) \cdot 10^{4}$ & $7.326(1) \cdot 10^{2}$ & $6.5(8) \cdot 10^{2}$ & $8.56(1) \cdot 10^{3}$ \\
$5.30884 \cdot 10^{-3}$ & $3.6440(3) \cdot 10^{1}$ & $-6.354(5) \cdot 10^{2}$ & $-1.79(4) \cdot 10^{4}$ & $7.034(1) \cdot 10^{2}$ & $1.13(8) \cdot 10^{3}$ & $8.04(1) \cdot 10^{3}$ \\
$5.95662 \cdot 10^{-3}$ & $3.5706(3) \cdot 10^{1}$ & $-5.963(5) \cdot 10^{2}$ & $-1.71(4) \cdot 10^{4}$ & $6.748(1) \cdot 10^{2}$ & $1.58(7) \cdot 10^{3}$ & $7.55(1) \cdot 10^{3}$ \\
$6.68344 \cdot 10^{-3}$ & $3.4974(3) \cdot 10^{1}$ & $-5.585(5) \cdot 10^{2}$ & $-1.67(3) \cdot 10^{4}$ & $6.467(1) \cdot 10^{2}$ & $2.04(7) \cdot 10^{3}$ & $7.06(1) \cdot 10^{3}$ \\
\hline
\end{tabular}
}
\caption{\label{table2_aachen}
Perturbative coefficients for the Aachen jet rates (continued).
}
\end{center}
\end{table}
\clearpage
\begin{table}[p]
\begin{center}
{\scriptsize
\begin{tabular}{|c|c|c|c|c|c|c|}
\hline
 $y_{cut}$ & $A_3$ & $B_3$ & $C_3$ & $B_4$ & $C_4$ & $C_5$ \\
\hline 
$1.05925 \cdot 10^{-5}$ & $8.0409(6) \cdot 10^{1}$ & $-5.613(7) \cdot 10^{3}$ & $-3.0(5) \cdot 10^{4}$ & $3.7882(9) \cdot 10^{3}$ & $-2.04(1) \cdot 10^{5}$ & $1.054(2) \cdot 10^{5}$ \\
$1.18850 \cdot 10^{-5}$ & $7.9682(6) \cdot 10^{1}$ & $-5.531(7) \cdot 10^{3}$ & $-3.0(5) \cdot 10^{4}$ & $3.7228(9) \cdot 10^{3}$ & $-2.02(1) \cdot 10^{5}$ & $1.029(2) \cdot 10^{5}$ \\
$1.33352 \cdot 10^{-5}$ & $7.8954(6) \cdot 10^{1}$ & $-5.440(6) \cdot 10^{3}$ & $-2.7(5) \cdot 10^{4}$ & $3.6583(9) \cdot 10^{3}$ & $-1.98(1) \cdot 10^{5}$ & $1.003(2) \cdot 10^{5}$ \\
$1.49624 \cdot 10^{-5}$ & $7.8228(6) \cdot 10^{1}$ & $-5.358(6) \cdot 10^{3}$ & $-3.1(4) \cdot 10^{4}$ & $3.5944(9) \cdot 10^{3}$ & $-1.93(1) \cdot 10^{5}$ & $9.78(2) \cdot 10^{4}$ \\
$1.67880 \cdot 10^{-5}$ & $7.7501(6) \cdot 10^{1}$ & $-5.270(6) \cdot 10^{3}$ & $-2.9(4) \cdot 10^{4}$ & $3.5310(9) \cdot 10^{3}$ & $-1.89(1) \cdot 10^{5}$ & $9.54(2) \cdot 10^{4}$ \\
$1.88365 \cdot 10^{-5}$ & $7.6775(6) \cdot 10^{1}$ & $-5.175(6) \cdot 10^{3}$ & $-2.7(4) \cdot 10^{4}$ & $3.4682(8) \cdot 10^{3}$ & $-1.85(1) \cdot 10^{5}$ & $9.29(2) \cdot 10^{4}$ \\
$2.11349 \cdot 10^{-5}$ & $7.6048(6) \cdot 10^{1}$ & $-5.084(5) \cdot 10^{3}$ & $-2.3(4) \cdot 10^{4}$ & $3.4059(8) \cdot 10^{3}$ & $-1.79(1) \cdot 10^{5}$ & $9.06(2) \cdot 10^{4}$ \\
\hline
\end{tabular}
}
\caption{\label{table1_antikt}
Perturbative coefficients for the anti-$k_t$ jet rates.
}
\end{center}
\end{table}
\begin{table}[p]
\begin{center}
{\scriptsize
\begin{tabular}{|c|c|c|c|c|c|c|}
\hline
 $y_{cut}$ & $A_3$ & $B_3$ & $C_3$ & $B_4$ & $C_4$ & $C_5$ \\
\hline 
$3.34965 \cdot 10^{-3}$ & $4.4024(3) \cdot 10^{1}$ & $-1.629(1) \cdot 10^{3}$ & $-1.20(7) \cdot 10^{4}$ & $1.2202(3) \cdot 10^{3}$ & $-3.34(2) \cdot 10^{4}$ & $2.140(3) \cdot 10^{4}$ \\
$3.75837 \cdot 10^{-3}$ & $4.3295(3) \cdot 10^{1}$ & $-1.570(1) \cdot 10^{3}$ & $-1.22(7) \cdot 10^{4}$ & $1.1829(3) \cdot 10^{3}$ & $-3.15(2) \cdot 10^{4}$ & $2.049(3) \cdot 10^{4}$ \\
$4.21696 \cdot 10^{-3}$ & $4.2566(3) \cdot 10^{1}$ & $-1.5135(9) \cdot 10^{3}$ & $-1.18(6) \cdot 10^{4}$ & $1.1464(2) \cdot 10^{3}$ & $-2.98(2) \cdot 10^{4}$ & $1.960(3) \cdot 10^{4}$ \\
$4.73151 \cdot 10^{-3}$ & $4.1836(3) \cdot 10^{1}$ & $-1.4568(9) \cdot 10^{3}$ & $-1.12(6) \cdot 10^{4}$ & $1.1104(2) \cdot 10^{3}$ & $-2.81(2) \cdot 10^{4}$ & $1.873(3) \cdot 10^{4}$ \\
$5.30884 \cdot 10^{-3}$ & $4.1108(3) \cdot 10^{1}$ & $-1.4020(8) \cdot 10^{3}$ & $-1.14(6) \cdot 10^{4}$ & $1.0750(2) \cdot 10^{3}$ & $-2.65(2) \cdot 10^{4}$ & $1.789(3) \cdot 10^{4}$ \\
$5.95662 \cdot 10^{-3}$ & $4.0377(3) \cdot 10^{1}$ & $-1.3488(8) \cdot 10^{3}$ & $-1.12(6) \cdot 10^{4}$ & $1.0401(2) \cdot 10^{3}$ & $-2.49(1) \cdot 10^{4}$ & $1.710(3) \cdot 10^{4}$ \\
$6.68344 \cdot 10^{-3}$ & $3.9646(3) \cdot 10^{1}$ & $-1.2954(8) \cdot 10^{3}$ & $-1.11(5) \cdot 10^{4}$ & $1.0057(2) \cdot 10^{3}$ & $-2.35(1) \cdot 10^{4}$ & $1.627(2) \cdot 10^{4}$ \\
\hline
\end{tabular}
}
\caption{\label{table2_antikt}
Perturbative coefficients for the anti-$k_t$ jet rates (continued).
}
\end{center}
\end{table}
\clearpage
\begin{table}[p]
\begin{center}
{\scriptsize
\begin{tabular}{|c|c|c|c|c|c|c|}
\hline
 $y_{cut}$ & $A_3$ & $B_3$ & $C_3$ & $B_4$ & $C_4$ & $C_5$ \\
\hline 
$1.05925 \cdot 10^{-5}$ & $7.5472(6) \cdot 10^{1}$ & $-4.040(5) \cdot 10^{3}$ & $-1.9(3) \cdot 10^{4}$ & $3.0027(7) \cdot 10^{3}$ & $-9.98(8) \cdot 10^{4}$ & $6.85(1) \cdot 10^{4}$ \\
$1.18850 \cdot 10^{-5}$ & $7.4745(5) \cdot 10^{1}$ & $-3.955(5) \cdot 10^{3}$ & $-1.8(3) \cdot 10^{4}$ & $2.9445(7) \cdot 10^{3}$ & $-9.67(8) \cdot 10^{4}$ & $6.66(1) \cdot 10^{4}$ \\
$1.33352 \cdot 10^{-5}$ & $7.4018(5) \cdot 10^{1}$ & $-3.876(5) \cdot 10^{3}$ & $-1.8(3) \cdot 10^{4}$ & $2.8871(7) \cdot 10^{3}$ & $-9.34(7) \cdot 10^{4}$ & $6.47(1) \cdot 10^{4}$ \\
$1.49624 \cdot 10^{-5}$ & $7.3292(5) \cdot 10^{1}$ & $-3.790(5) \cdot 10^{3}$ & $-2.1(3) \cdot 10^{4}$ & $2.8299(7) \cdot 10^{3}$ & $-8.93(7) \cdot 10^{4}$ & $6.29(1) \cdot 10^{4}$ \\
$1.67880 \cdot 10^{-5}$ & $7.2565(5) \cdot 10^{1}$ & $-3.703(5) \cdot 10^{3}$ & $-1.9(3) \cdot 10^{4}$ & $2.7736(7) \cdot 10^{3}$ & $-8.58(7) \cdot 10^{4}$ & $6.10(1) \cdot 10^{4}$ \\
$1.88365 \cdot 10^{-5}$ & $7.1838(5) \cdot 10^{1}$ & $-3.623(4) \cdot 10^{3}$ & $-2.2(3) \cdot 10^{4}$ & $2.7177(7) \cdot 10^{3}$ & $-8.24(7) \cdot 10^{4}$ & $5.92(1) \cdot 10^{4}$ \\
$2.11349 \cdot 10^{-5}$ & $7.1110(5) \cdot 10^{1}$ & $-3.538(4) \cdot 10^{3}$ & $-2.3(3) \cdot 10^{4}$ & $2.6625(6) \cdot 10^{3}$ & $-7.93(6) \cdot 10^{4}$ & $5.74(1) \cdot 10^{4}$ \\
\hline
\end{tabular}
}
\caption{\label{table1_siscone}
Perturbative coefficients for the SISCone jet rates.
}
\end{center}
\end{table}
\begin{table}[p]
\begin{center}
{\scriptsize
\begin{tabular}{|c|c|c|c|c|c|c|}
\hline
 $y_{cut}$ & $A_3$ & $B_3$ & $C_3$ & $B_4$ & $C_4$ & $C_5$ \\
\hline 
$3.34965 \cdot 10^{-3}$ & $3.9087(3) \cdot 10^{1}$ & $-7.339(7) \cdot 10^{2}$ & $-1.84(5) \cdot 10^{4}$ & $7.855(2) \cdot 10^{2}$ & $5.2(9) \cdot 10^{2}$ & $9.19(1) \cdot 10^{3}$ \\
$3.75837 \cdot 10^{-3}$ & $3.8357(3) \cdot 10^{1}$ & $-6.922(7) \cdot 10^{2}$ & $-1.82(4) \cdot 10^{4}$ & $7.553(1) \cdot 10^{2}$ & $1.05(9) \cdot 10^{3}$ & $8.64(1) \cdot 10^{3}$ \\
$4.21696 \cdot 10^{-3}$ & $3.7628(3) \cdot 10^{1}$ & $-6.512(7) \cdot 10^{2}$ & $-1.77(4) \cdot 10^{4}$ & $7.258(1) \cdot 10^{2}$ & $1.65(8) \cdot 10^{3}$ & $8.12(1) \cdot 10^{3}$ \\
$4.73151 \cdot 10^{-3}$ & $3.6898(3) \cdot 10^{1}$ & $-6.109(6) \cdot 10^{2}$ & $-1.67(4) \cdot 10^{4}$ & $6.968(1) \cdot 10^{2}$ & $2.21(8) \cdot 10^{3}$ & $7.62(1) \cdot 10^{3}$ \\
$5.30884 \cdot 10^{-3}$ & $3.6168(3) \cdot 10^{1}$ & $-5.715(6) \cdot 10^{2}$ & $-1.62(4) \cdot 10^{4}$ & $6.683(1) \cdot 10^{2}$ & $2.68(8) \cdot 10^{3}$ & $7.14(1) \cdot 10^{3}$ \\
$5.95662 \cdot 10^{-3}$ & $3.5439(3) \cdot 10^{1}$ & $-5.335(6) \cdot 10^{2}$ & $-1.61(4) \cdot 10^{4}$ & $6.404(1) \cdot 10^{2}$ & $3.15(7) \cdot 10^{3}$ & $6.68(1) \cdot 10^{3}$ \\
$6.68344 \cdot 10^{-3}$ & $3.4707(3) \cdot 10^{1}$ & $-4.971(5) \cdot 10^{2}$ & $-1.60(4) \cdot 10^{4}$ & $6.131(1) \cdot 10^{2}$ & $3.44(7) \cdot 10^{3}$ & $6.23(1) \cdot 10^{3}$ \\
\hline
\end{tabular}
}
\caption{\label{table2_siscone}
Perturbative coefficients for the SISCone jet rates (continued).
}
\end{center}
\end{table}

\clearpage

\end{document}